\documentclass[11pt,a4paper]{article}
\usepackage{mathtools}
\usepackage{graphicx}
\usepackage{amsmath}
\usepackage{epsf,amsfonts,amssymb,epsfig,color,fancyhdr,lastpage}
\usepackage[a4paper,left=2.5cm,right=2.5cm,top=2.5cm,bottom=2.5cm]{geometry}
\usepackage[colorlinks=true,linktocpage=true,linkcolor=blue,citecolor=blue]{hyperref}
\usepackage[labelformat=simple]{subcaption}

\newcommand{\dd}{\mathrm{d}}
\newcommand{\kappaHP}{\tilde{\kappa}_{\mathrm{HP}}}
\newcommand{\rH}{\tilde{r}_{\hspace{-0.1em}H}}
\newcommand{\lleft}{\mathopen{}\left(\mathclose{}}
\newcommand{\rright}{\mathopen{}\right)\mathclose{}}
\DeclareMathOperator{\trace}{\mathrm{Tr}}
\numberwithin{equation}{section}
\title{\LARGE \bf Black rings in global anti-de Sitter space}

\author{Pau Figueras and Saran Tunyasuvunakool}

\date{}

\begin{document}

\maketitle

\thispagestyle{empty}

\begin{center}
DAMTP, Centre for Mathematical Sciences, \\
\vskip .25cm
Wilberforce Road, Cambridge CB3 0WA, U.K.
\\
\vskip .5cm
\texttt{p.figueras@damtp.cam.ac.uk, s.tunyasuvunakool@damtp.cam.ac.uk}
\end{center}

\vskip 1cm
\begin{abstract}
We construct five dimensional black rings in global anti-de Sitter space using numerical methods. These rings satisfy the BPS bound $|J| < M\,\ell$, but the angular velocity always violates the Hawking--Reall bound $\left| \Omega_H\,\ell \right| \leq 1$, indicating that they should be unstable under superradiance. At high temperatures, the limit $\left| \Omega_H\,\ell \right| \searrow 1$ is attained by thin rings with an arbitrarily large radius. However, at sufficiently low temperatures, this limit is saturated by a new kind of rings, whose outer circle can still be arbitrarily long while the hole in the middle does not grow proportionally. This gives rise to a membrane-like horizon geometry, which does not have an asymptotically flat counterpart. We find no evidence for thin AdS black rings whose transverse $S^2$ is much larger than the radius of AdS, $\ell$, and thus these solutions never fall into the hydrodynamic regime of the dual CFT. Thermodynamically, we find that AdS black rings never dominate the grand canonical ensemble. The behaviour of our solutions in the microcanonical ensemble approaches known perturbative results in the thin-ring limit. 
\end{abstract}

\newpage
\setcounter{page}{1}

%%%%%%%%%%%%%%%%%%%%%%%%%%%%%%%%%%%%%%%%%%%%%%%
\section{Introduction and summary of the main results}
\label{sec:intro}
%%%%%%%%%%%%%%%%%%%%%%%%%%%%%%%%%%%%%%%%%%%%%%%
The study of equilibrium black holes in $D\geq 5$ (see \cite{Emparan:2008eg} for a review) has revealed that the physics of these objects can be very different from that of their four-dimensional counterparts. In particular, black holes with a non-spherical horizon topology, such as black rings \cite{Emparan:2001wn,Pomeransky:2006bd,Emparan:2006mm}, black ringoids \cite{Kleihaus:2014pha}, and regular multi-black hole spacetimes in vacuum \cite{Elvang:2007rd,Iguchi:2007is,Evslin:2007fv,Izumi:2007qx,Elvang:2007hs}, among others \cite{Emparan:2009vd,Kunduri:2010vg,Dias:2014cia,Emparan:2014pra}, have been shown to exist. These types of black objects do not exist in $D=4$, and thus they possess fundamentally new physical properties. A natural question to ask is whether these objects can also exist in asymptotically anti-de Sitter (AdS) space. This question is further motivated by the gauge/gravity correspondence \cite{Maldacena:1997re}, according to which these new types of stationary black holes should correspond to new finite temperature phases of gauge theories. Furthermore, AdS is a maximally symmetric space and, as such, can be regarded from a mathematical viewpoint as being as fundamental as the Minkowski space. The study of black holes in AdS is therefore also an interesting question in its own right. 

Through the AdS/CFT duality, one can look for new types of black holes by solving the equations of motion of the dual field in their \emph{hydrodynamic regime}, in which they simplify considerably. Using this approach, \cite{Bhattacharyya:2007vs} looked for solutions to the Navier--Stokes equations on the Einstein static universe, $\mathbb R_t\times S^3$, corresponding to rotating fluid configurations. However, they found only stationary fluid configurations which are dual rotating spherical black holes in AdS which are already known \cite{Carter:1968ks,Hawking:1998kw,Gibbons:2004uw}. Still, the assumption of AdS asymptotics also allows for more general boundary conditions, such as Scherk--Schwarz compactifications of AdS. In these settings, \cite{Lahiri:2007ae} constructed solutions to the relativistic Navier--Stokes equations corresponding to rotating plasma balls and plasma rings, and hence they were able to study the phase diagram of rotating black holes in such spacetimes. Of course, this approach can only capture the physics of black holes that admit a hydrodynamic limit, which unfortunately is not always the case.   

In this paper we will focus on the study of stationary black rings in global AdS. For simplicity, we will concentrate only on the $D=5$ case, which is more interesting from the point of view of the gauge/gravity correspondence, and leave the higher dimensional cases for future work. Therefore, the horizon topology of the black holes that we consider in this paper is $S^1 \times S^2$.  The solution generating techniques \cite{Belinsky:1979mh,Emparan:2001wk,Harmark:2004rm} that have been so successfully used in the studies of 5D asymptotically flat (AF) stationary vacuum black holes do not seem to have a straightforward extension to the AdS case. In spite of this, Ref. \cite{Caldarelli:2008pz} was able to access black rings in global AdS for the first time using approximate methods. 

In AF space, black rings come in two families, namely the \emph{thin} and \emph{fat} ones, depending on the ratio between the typical sizes of the horizon $S^1$, denoted $R_{S^1}$, and of the horizon $S^2$, denoted $R_{S^2}$. In AdS, however, there is yet another length scale which can play a role in the physics of black rings, namely the radius of AdS, denoted $\ell$. In order to take the latter into account,  Ref. \cite{Caldarelli:2008pz} introduced the following terminology to describe various notions of size for black rings in AdS:
\begin{itemize}
\item \textit{Thin} rings have $R_{S^2}\ll R_{S^1}$, while \textit{fat} rings have $R_{S^2}\gtrsim R_{S^1}$.
\item \textit{Small} rings have $R_{S^2}<\ell$, while $\textit{large}$ rings have $R_{S^2} > \ell$.
\item \textit{Short} rings have $R_{S^1}<\ell$, while $\textit{long}$ rings have $R_{S^1}>\ell$.
\end{itemize}
According to this terminology, the approximation of \cite{Caldarelli:2008pz} is valid for small thin rings, which can be either short or long.\footnote{Note that Ref. \cite{Caldarelli:2008pz} did not assume any hierarchy between the radius of the $S^1$ of the ring and $\ell$.} Based on quantitative results for thin rings  and some educated guesses, \cite{Caldarelli:2008pz} put forward a proposal for the phase diagram of black rings in AdS in the microcanonical ensemble. According to this proposal, the phase diagram of black rings in AdS is qualitatively similar to the phase digram in the asymptotically flat case, but it is compressed into the range $J\leq M\,\ell$. In this paper, we use numerical methods to construct black rings in global AdS which can be thin or fat, small or large, and short or long. This allows us complete the phase diagram of 5D black rings in AdS. In particular,  we confirm that fat rings merge with the spherical black holes at a singular solution with zero area, as in the AF case. This suggests that, in $D\geq 6$, the pattern of connections between various stationary black hole phases of different topologies conjectured in \cite{Caldarelli:2008pz} is indeed correct. One of the main unanswered questions in \cite{Caldarelli:2008pz} was whether rings which are both thin and large exist.  We address this question in this paper and, quite confidently, we find no evidence for such thin large rings. 

We use the method of \cite{Headrick:2009pv,Adam:2011dn} to construct black rings in AdS numerically. In this approach, it is more natural to study stationary solutions of the Einstein equations in the grand canonical ensemble, since the temperature of the black hole, $T_H$, and its angular velocity, $\Omega_H$, naturally appear as directly specifiable boundary data. Therefore, before describing our results for black rings, it is worth reviewing the grand canonical ensemble for rotating spherical black holes in AdS. As is well-known, in the static limit, $\Omega_H=0$, black holes only exist for temperatures greater than the Hawking--Page temperature, $T_\mathrm{HP} \coloneqq \frac{\sqrt{2}}{\pi\,\ell}$ \cite{Hawking:1982dh}. Furthermore, static black holes in AdS are classified as \emph{large} or \emph{small} according to their size compared to the radius of AdS. On the other hand, in the rotating case, $\Omega_H\neq 0$, black holes can exist for any non-zero temperature, but it no longer makes sense to distinguish between ``large'' and ``small'' black holes from a geometrical point of view. One can, however, still classify black holes according to their thermodynamical stability.\footnote{We want to remind the reader that the thermodynamic stability of black holes does not necessarily have any correlation to their actual dynamical stability.} In analogy to the static case,  we shall refer to the thermodynamically stable black holes as ``large'' black holes, and to the thermodynamically unstable ones as ``small'' black holes. We must emphasise that this terminology does not necessarily reflect the geometric size of the black hole. Large rotating black holes only exist for temperatures greater than some critical temperature,  $T_H > T_c \coloneqq \frac{1}{\pi\,\ell}$, whilst small rotating black holes exist at all non-zero temperatures. Moreover, large black holes always obey the Hawking--Reall bound, $|\Omega_H\,\ell|<1$ \cite{Hawking:1999dp}.  Rotating AdS black holes satisfying this bound admit a globally defined timelike Killing vector field, and this implies that they should also be classically dynamically stable \cite{Hawking:1999dp}. On the other hand, some small rotating black holes violate this bound, and hence one deduces that they could be classically dynamically unstable, in particular under the superradiant instability. The latter has been recently studied in detail in \cite{Cardoso:2013pza}.

\begin{figure}[tp]
\begin{center}
\begin{subfigure}[b]{0.5\linewidth}
    \centering
    \includegraphics[width=8.2cm]{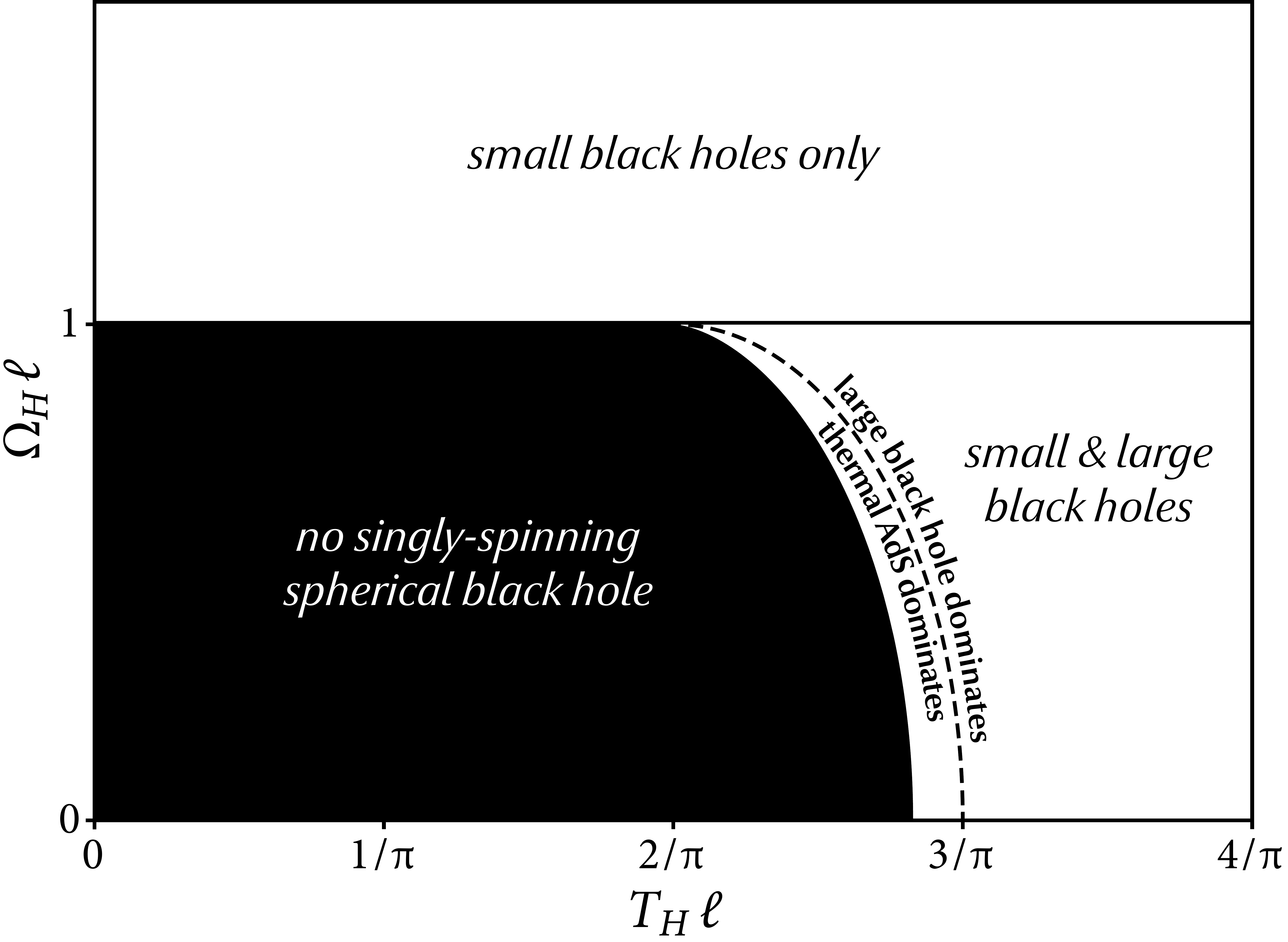}
    \caption{spherical AdS black holes}
    \label{subfig:phasesSph}
\end{subfigure}%
\begin{subfigure}[b]{0.5\linewidth}
    \centering
    \includegraphics[width=8.2cm]{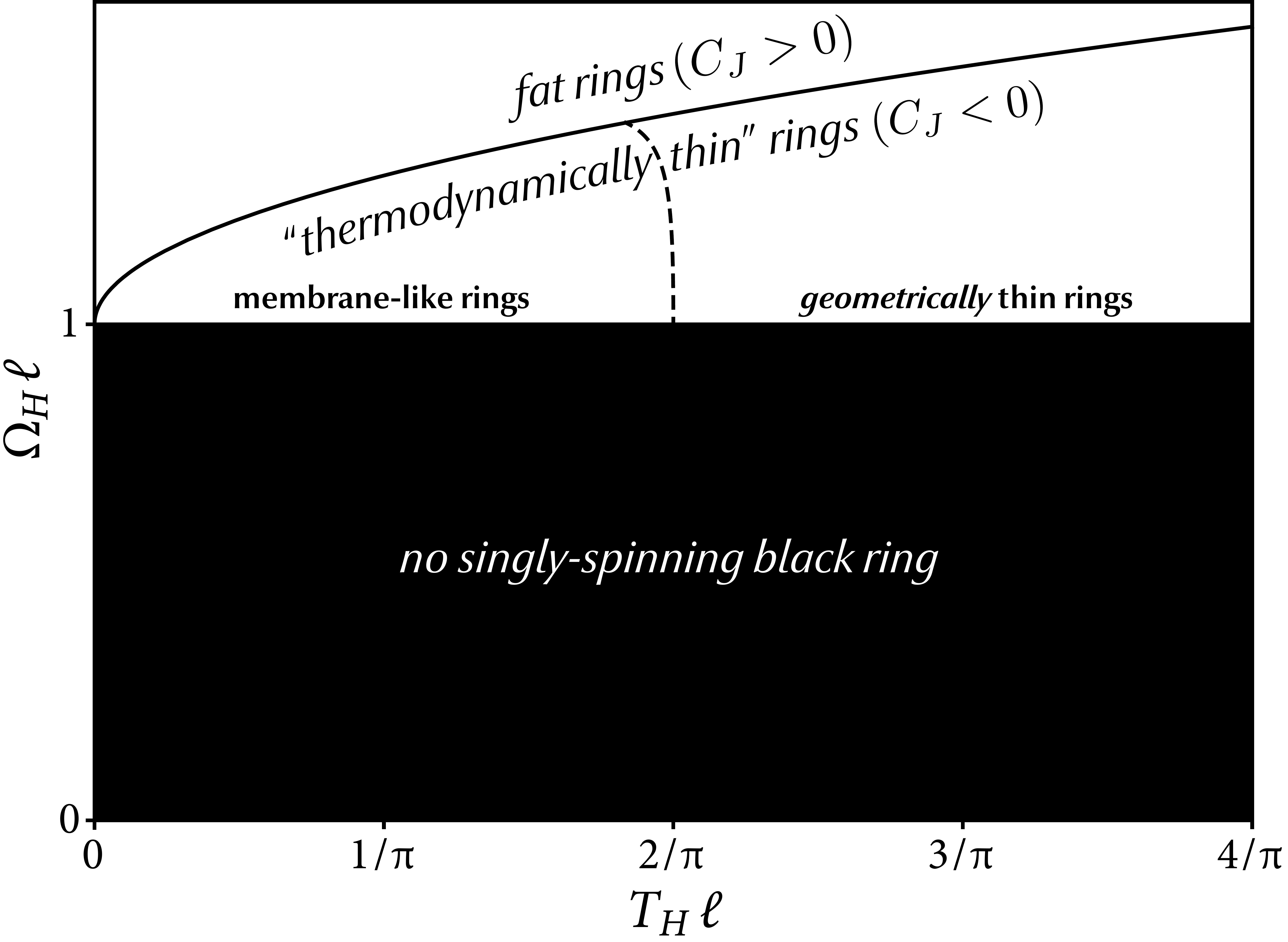}
    \caption{AdS black rings}
    \label{subfig:phasesRing}
\end{subfigure}
\end{center}
\caption{Phase diagrams of singly-spinning \subref{subfig:phasesSph} spherical black holes and \subref{subfig:phasesRing} black rings in 5D global AdS, presented in the $(T_H, \Omega_H)$ plane. Both kinds of objects exists at all non-zero temperatures. All of our AdS black rings have $|\Omega_H\,\ell|>1$. The black rings can be classified into two distinct phases according to their heat capacity at constant angular momentum, $C_J$. Those with $C_J > 0$ are \emph{fat rings}, as in the AF case, while those with $C_J < 0$ exhibit somewhat more complicated behaviour. At high temperatures, $T_H\,\ell > 1/\pi$, these rings are indeed \emph{thin}, again as in the AF case. However, for $T_H\,\ell < 1/\pi$ the outer radius of the $S^1$ these rings grows much more quickly than its inner radius as $\Omega_H\,\ell \rightarrow 1$, causing the $S^2$ to become highly stretched. In this regime, the ring no longer becomes ``thin'' in the geometric sense, but instead approaches a membrane-like geometry.}
\label{fig:phases}
\end{figure}

In this paper we place AdS black rings among the known stationary black hole phases in the grand canonical ensemble. One of our main results is that black rings in AdS \textit{never} dominate the grand canonical ensemble. Moreover, AdS black rings always obey $|\Omega_H\,\ell|>1$, and hence they should \emph{all}  be classically dynamically unstable under superradiance. The thermodynamic behaviour of AdS rings is qualitatively and quantitatively similar to that of small rotating black holes. The phase diagram for AdS black rings, expressed in terms of $T_H$ and $\Omega_H$, can be summarised as follows:
\begin{itemize}
\item $T_H>T_c$: the $|\Omega_H\,\ell| \to \infty$ limit is reached from the fat family of rings, and the limit is saturated by a singular solution which merges with the small spherical black hole family, as in the AF case. On the other hand, the $|\Omega_H\,\ell| \to 1$ limit corresponds to an infinitely thin and long ring, and hence accessible using the perturbation theory of \cite{Caldarelli:2008pz}. 
\item $T^*<T_H<T_c$: the limit $|\Omega_H\,\ell| \to \infty$ still corresponds to the (singular) merger of a fat ring with the spherical black hole. However, the $\Omega_H\,\ell \to 1$ limit now corresponds to a new membrane-like limit for rings which are \emph{not} geometrically thin. In this limit, black rings tend to the same singular black membrane-type solution, with horizon topology $\mathbb H^2\times S^1$, as the small black holes (see \cite{Caldarelli:2008pz}). For intermediate angular velocities, geometrically thin rings can still occur. 
\item $T_H<T^*$: there are no geometrically thin rings below a certain temperature $T^*$. This may be related to the fact that we find no evidence for long thin rings which are also large. The limits $|\Omega_H\,\ell| \to \infty$  and $|\Omega_H\,\ell| \to 1$ respectively are reached by fat and membrane-like rings respective, as in the previous case. We find that $T^*\ell > \frac{1}{2\,\pi}$.
\end{itemize}
We sketch the various black ring phases on the $(T_H,\Omega_H)$ plane in Fig. \ref{fig:phases}. Finally, we note that neither fat rings nor membrane-like rings seem to develop new negative modes of the Lichnerowicz operator as one moves along the family of solutions. This suggest that, at least in 5D, these black rings are not unstable under a Gregory--Laflamme type of instability along the rotation plane. 

 Using standard holographic renormalisation techniques \cite{deHaro:2000xn}, we extract the one-point function for the stress-energy tensor of the dual $\mathcal N=4$ super Yang-Mills (SYM) field on $S^3$. This observable allows us to get some idea as to how the dual field theory might be affected by the horizon topology of the black hole in the bulk.  We find that long thin rings can get very close to the boundary of AdS, giving rise to an energy distribution which is highly localised on one of the poles of the boundary $S^3$. This energy distribution is very different from the one corresponding to the small black hole with the same temperature and angular velocity. Unsurprisingly, fat rings give rise to an energy distribution which is qualitatively and quantitatively very similar to that of the corresponding small black hole. It would be interesting to investigate other observables, such as higher point correlation functions,  Wilson loops or entanglement entropy. In Fig. \ref{fig:Edensity} we depict the energy distribution of some representative black rings. 
  
 The rest of the paper is organised as follows. In \S\ref{sec:setup} we describe our numerical construction of black rings in AdS and in \S\ref{sec:physQ} we explain how we calculate various physical quantities, in particular the mass, from our numerical solutions. In \S\ref{sec:geometry}, we study the geometry of the horizon of the AdS black rings, and in \S\ref{sec:thermo} we study the thermodynamics and produce phase diagrams for the grand canonical and microcanonical ensembles. Finally, in \S\ref{sec:stresstensor}, we extract the v.e.v. of the stress tensors of the dual CFT and compare them with that of the spherical black holes.

Throughout this paper, we shall use tilde to denote quantities which have been nondimensionalised with respect to the AdS radius $\ell$. In particular, we make the following definitions
\begin{equation}
\tilde{M} \coloneqq M \, \ell^{-2} \quad,\quad 
\tilde{\kappa} \coloneqq \kappa \, \ell \quad,\quad \tilde{A}_{H} \coloneqq A_{H} \, \ell^{-3} \quad,\quad
\tilde{\Omega}_{H} \coloneqq \Omega_{H} \, \ell \quad,\quad \tilde{J} \coloneqq J \, \ell^{-3} .
\end{equation}

%%%%%%%%%%%%%%%%%%%%%%%%%%%%%%%%%%%%%%%%%%%%%%%
\section{Numerical construction of AdS black rings}
\label{sec:setup}
%%%%%%%%%%%%%%%%%%%%%%%%%%%%%%%%%%%%%%%%%%%%%%%
In this section, we explain the methods that we use to numerically construct black rings in global AdS. We begin in \S\ref{subsec:setup} by discussing our choice of coordinates and the formulation of our spacetime metric ansatz. We then proceed to explain our construction of the reference metric in \S\ref{subsec:gauge}, followed by the description of our boundary conditions in \S\ref{subsec:BCs}. We finish the section with further technical details in \S\ref{subsec:code}

\subsection{Basic setup}
\label{subsec:setup}

Our goal is to construct black rings in 5D with the asymptotics of global AdS. These are stationary black hole solutions to the vacuum Einstein equations with a negative cosmological constant, whose spatial horizon topology is $S^1\times S^2$. For simplicity, we shall restrict ourselves to \emph{singly-spinning} black rings, i.e. those which are only rotating along the $S^1$ direction, while leaving the doubly-spinning case for future work. Therefore,  our solutions should be thought of as the AdS-generalisation of the black rings of \cite{Emparan:2001wn}.\footnote{Ref. \cite{Figueras:2005zp} constructed an AF black ring with rotation only on the $S^2$ but this solution has conical singularities. One should expect that the generalisation of this solution to global AdS should also suffer from conical singularities, but we will not investigate this here.} 

The isometry group of the AdS rings is the same as that of the AF rings, namely $\mathbb R_{t}\times U(1)^2$, corresponding to time translations and rotations on the two 2-planes. In the AF case, the Einstein equations restricted to spacetimes with this isometry group are completely integrable. This makes it possible to explicitly construct all solutions within this class using an algebraic procedure \cite{Belinsky:1979mh,Emparan:2001wk,Harmark:2004rm}. Unfortunately, it is not currently known whether this integrability persists in the presence of a cosmological constant. Therefore, in this paper we shall rely on numerical methods to construct our solutions. Specifically, we numerically solve the Einstein--DeTurck equations on our spacetime $(\mathcal M, g)$, 
\begin{equation}
R_{ab}-\nabla_{(a}\xi_{b)} + \frac{4}{\ell^2}=0\,, \quad \textrm{with} \quad \xi^a \coloneqq g^{bc}\left(\Gamma^a_{\phantom a bc}-\bar\Gamma^a_{\phantom a bc}\right)\,,
\label{eq:EDT}
\end{equation}
where $g$ is the physical spacetime metric that we seek, $\Gamma$ is its corresponding Levi-Civita connection, and $\bar\Gamma$ is the Levi-Civita connection associated to a reference metric $\bar g$ on the spacetime manifold $\mathcal M$.  This method has now become standard in the field and we refer the reader to the references \cite{Headrick:2009pv,Adam:2011dn,Wiseman:2011by} for more details. Under certain regularity assumptions, \cite{Figueras:2011va} proved that all \textit{static} solutions to the Einstein--DeTurck equations are in fact Einstein. A similar result is not yet available for stationary spacetimes, and therefore we must check \textit{a posteriori} that our solutions are Einstein to within a numerical tolerance. This is indeed the case for the solutions presented in this paper.

Ref. \cite{Emparan:2001wn} (see also \cite{Emparan:2004wy,Emparan:2006mm}) wrote down the black ring metric in a $C$-metric type of coordinates. These coordinates are adapted to the geometry of the black ring in the sense that they foliate the spatial slices of the spacetime with surfaces of ring-like topology. Similar coordinate system can be used in AdS \cite{Dias:2002mi}, but so far they have not facilitated the finding of the AdS black ring metric analytically. The drawback is that this coordinate system is singular at infinity, in the sense that the latter is represented by a single point. Whilst in principle one could use this type of coordinate system to numerically construct solutions, it would be essential to first analyse the non-trivial (singular) behaviour of the various metric coefficients near the boundary of AdS. Moreover, in the discretised system, spatial infinity would inevitably be poorly resolved, which could become an issue when it comes to extracting the stress tensor of the dual CFT. We will therefore follow a different strategy and use two coordinate systems to cover different parts of the spacetime: one coordinate system will be adapted to the \emph{``outer''} region, near the boundary of AdS, whilst the other coordinate system will be adapted to cover the \emph{``inner''} region, near the horizon of the ring. 

The geometry near the AdS black ring horizon can be thought of as a deformation of the near-horizon geometry of the AF black ring. In general, this deformation is not necessarily small, but at very high temperatures (i.e., small rings) we can expect the near-horizon geometry to be very close to that of the AF solution. Therefore, we build our inner region ansatz by ``dressing" the AF black ring metric:
\begin{equation}
\begin{aligned}
\dd s^2_\textrm{inner}=&-\mathcal{T}_0\,e^{\mathcal{T}}\,\dd t^2+\mathcal{X}_0\,e^{\mathcal{X}}\,\dd x^2+\mathcal{Y}_0\,e^{\mathcal{Y}}(\dd y-\mathcal{W}\,\dd x)^2 \\
&+\mathcal{U}_0\,e^{\mathcal{U}}\,\dd\phi^2+\mathcal{V}_0\,e^{\mathcal{V}} \left( \dd\psi-\mathcal{Z}_0 \left( 1 + \mathcal{Z} \right) \dd t \right)^2 \, ,
\end{aligned}
\label{eq:metricNear}
\end{equation}
where $\mathcal{F} \coloneqq \left( \mathcal{T}, \mathcal{X}, \mathcal{Y}, \mathcal{U}, \mathcal{V}, \mathcal{W}, \mathcal{Z} \right)$ are unknown functions of $(x,y)$, and $\mathcal{F}_0 \coloneqq \left( \mathcal{T}_0, \mathcal{X}_0, \mathcal{Y}_0, \mathcal{U}_0, \mathcal{V}_0, \mathcal{Z}_0 \right)$ are functions which are analytically prescribed so that we recover the AF black ring line element when $\mathcal{F} \equiv 0$. In order to normalise the coordinate ranges and impose boundary conditions, we found it useful to transform the ring-like $(x,y)$ coordinates from the ones described in \cite{Emparan:2004wy,Emparan:2006mm} to 
\begin{equation}
x \to \cos(\pi\,x)\,,\qquad y \to -\frac{1+\nu+(1-y)\cos(\pi\,y)}{2\,\nu}\,.
\label{eq:XYcoords}
\end{equation}
The ranges of these transformed coordinates then become $0 \leq \{x, y \} \leq 1$ in this near horizon region. Here $\nu$ is the dimensionless parameter introduced in \cite{Emparan:2004wy}, which is related to the ring's surface gravity $\kappa$ and horizon angular velocity $\Omega_H$ by
\begin{equation}
\nu = \frac{\Omega_H}{\sqrt{4 \,\kappa^2 + \Omega_H^2}} .
\end{equation}
For the AF black ring the range of this parameter is $0<\nu<1$, however in AdS we found that the lower bound increases as the temperature lowers, see \S\ref{sec:canonical}. Writing down the AF ring in these new $(x,y)$ coordinates allows us to identify the expressions for the functions $\mathcal{F}_0$. 

Near the conformal boundary of the spacetime, we expect that the black ring spacetime will be just a small deformation of global AdS. Therefore, to cover the outer region, we use the most general metric which is manifestly asymptotically AdS and that is closed under diffeomorphisms which preserve the $\mathbb R_{t} \times U(1)^2$ isometry group:
\begin{equation}
\begin{aligned}
\dd s^2_\textrm{outer}=&-\left(1+\lambda^2\,R^2\right)\,e^{T}\,\dd t^2+\frac{e^{X}}{1+\lambda^2\,R^2}\,(\dd R-W\,\dd a)^2 \\
&+R^2 \left[\textstyle{\frac{\pi^2}{4}}\,e^{Y}\,\dd a^2+\cos^2 \mathopen{}\left(\mathclose{} \textstyle{\frac{\pi\,a}{2}} \mathopen{}\right)\mathclose{} \,e^{U}\,\dd\phi^2
+\sin^2 \mathopen{}\left(\mathclose{} \textstyle{\frac{\pi\,a}{2}} \mathopen{}\right)\mathclose{} \,e^{V} \left( \dd\psi-\mathcal{Z}_0 \left( 1 + Z \right) \dd t \right)^2 \right] ,
\end{aligned}
\label{eq:metricFar}
\end{equation}
where $\lambda \coloneqq \ell^{-1}$ is the inverse of the AdS radius, $F \coloneqq ( T,X,Y,U,V,W,Z )$ are unknown functions of $(R,a)$, and $\mathcal{Z}_0$ is identical to the function that appeared in (\ref{eq:metricNear}) but transformed into $(R,a)$ coordinates via a relation which we shall explain in due course. We use $\lambda$ rather than $\ell$ as a parameter in our solutions as it is computationally easier to make contact with the AF rings by setting $\lambda = 0$ rather than trying to make $\ell$ very large.

In order to cover the AdS boundary, we define a compact radial coordinate $r$ via
\begin{equation}
R = \frac{r}{1-r^2/k^2}\,,
\label{eq:compactification}
\end{equation}
where $k$ is some constant with dimensions of length. The ranges of these coordinates are ${r_\textrm{min} \leq r \leq k}$ and ${0 \leq a \leq 1}$, where $r_\textrm{min}$ is a parameter that sets the location where we switch between the outer and inner region patches. This is in principle an arbitrary gauge choice, as long as the coordinates remain well-defined everywhere (see below). In practice, we found that choosing $r_\textrm{min} \approx k / 2$ worked best. 

\begin{figure}[t]
\begin{center}
\includegraphics[scale=0.35]{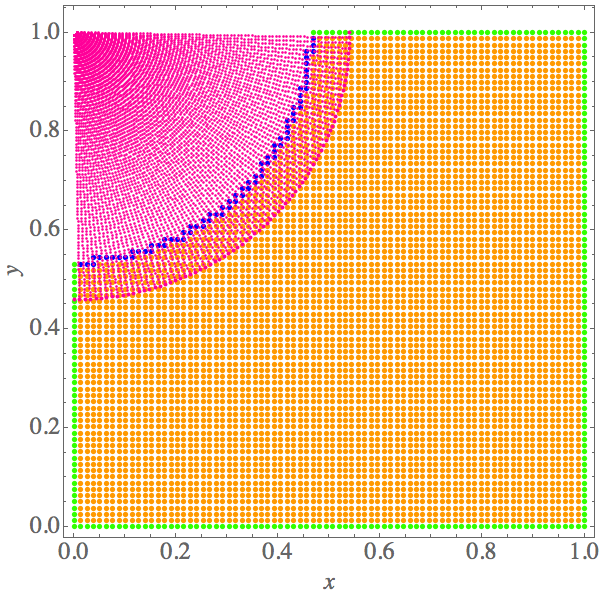}
\hspace{0.5cm}
\includegraphics[scale=0.35]{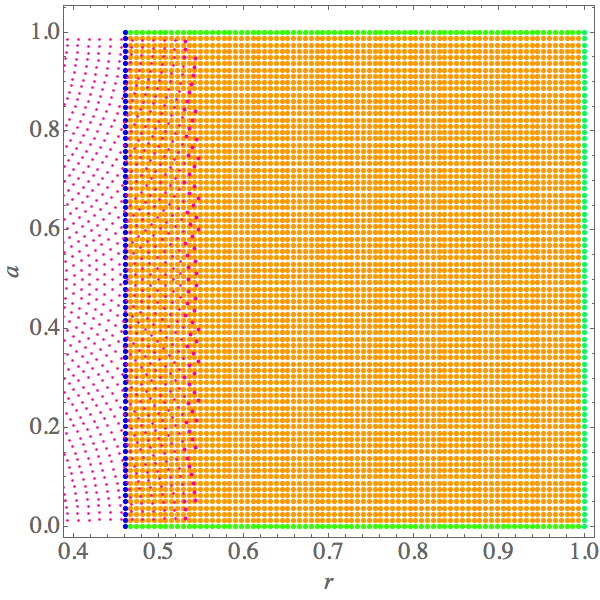}
\end{center}
\caption{Our computational grid, shown in the inner (left) and outer (right) region coordinates. Orange dots show grid points where the inner patch ansatz (\ref{eq:metricNear}) is used. Pink dots show grid points where the outer patch ansatz (\ref{eq:metricFar}) is used and vice versa. Green dots show grid points where boundary conditions are imposed. Blue dots show grid points where function values are obtained by interpolating data from the other patch. Note that these diagrams are only illustrative as the grid resolution used for actual calculations is significantly higher than what is shown above.}
\label{fig:patches}
\end{figure}

The outer region coordinates $(r,a)$ and the inner region ones $(x,y)$ are related by a simple coordinate transformation, 
\begin{equation}
x=(1-r/k)\,\cos\left(\textstyle{\frac{\pi\,a}{2}}\right)\,,\qquad y = 1-(1-r/k)\,\sin\left(\textstyle{\frac{\pi\,a}{2}}\right)\,.
\end{equation}
Note that the coordinate change becomes singular when $r=0$, which sets a lower bound on the parameter $r_\textrm{min}$ introduced above. Our choice avoids any of such issues. We henceforth choose to fix $k=1$ throughout, thereby setting the scale for the solutions. We depict the two coordinate patches in Fig. \ref{fig:patches}.

\subsection{Reference metric}
\label{subsec:gauge}

\begin{figure}[t]
\begin{center}
\includegraphics[scale=0.55]{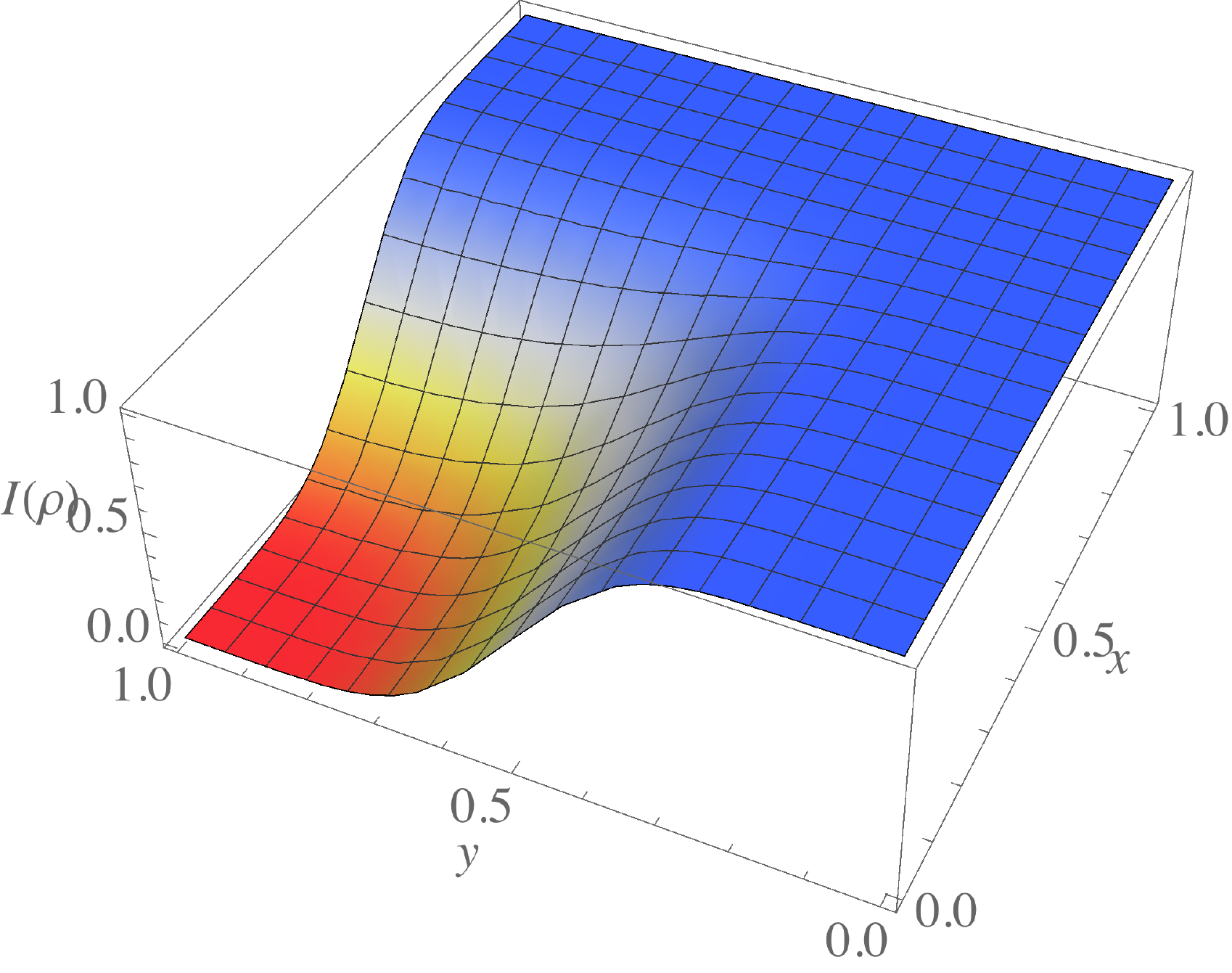}
\end{center}
\caption{Interpolating function $I(\hat{r})$ in terms of the inner region coordinates $(x,y)$. This is a smooth function defined everywhere but it is not compactly supported.}
\label{fig:interpolating}
\end{figure}

In the Einstein--DeTurck method, the gauge condition is determined simultaneously with the solution. The particular choice of gauge is made by fixing a reference metric $\bar g$ on the spacetime manifold $\mathcal M$. To do so, we follow \cite{Headrick:2009pv} and construct a suitable reference metric simply by considering
\begin{equation}
\bar g_{ab}\,dx^a\,dx^b = \left[ 1-I(\hat{r}) \right] \,\bar{g}^\textrm{outer}_{ab}\,dx^a\,dx^b+I(\hat{r})\, \bar{g}^\textrm{inner}_{ab}\,dx^a\,dx^b\,,
\label{eq:refmetric}
\end{equation}
where $\bar g^\textrm{outer}_{ab}$ is obtained from \eqref{eq:metricFar} and $\bar g^\textrm{inner}_{ab}$ is obtained from \eqref{eq:metricNear} after setting ${ \mathcal{F} \equiv 0 \equiv F }$. $I(\hat{r})$ is an interpolating function which is a function of a suitably defined coordinate $\hat{r}$ satisfying $\hat{r} = 0$ at the horizon and $\hat{r} = 1$ at infinity. The requirements on $I(\hat{r})$ are that it is smooth and monotonic\footnote{Strictly speaking we only need $0 \leq I(\hat{r}) \leq 1$ with equality only at the endpoints. However, we found that the system behave better when $I(\hat{r})$ is monotonic.}, with $I(0) = 1$ and $I(1) = 0$. In previous works, e.g. \cite{Headrick:2009pv}, one used an interpolating function whose compact support is limited to some subregion which does \emph{not} extend to the boundaries, thus manifestly ensuring that the reference metric satisfied all the boundary conditions. However, such a function tends to have large derivatives which are inevitably inherited by the solutions, and in our present setting we found that this proved problematic for the numerics. Instead, we use an interpolation function which \emph{is} supported near the boundaries, but whose normal derivatives vanish up to at least fourth order, both at the horizon and at the boundary of AdS. This still ensures that the reference metric \eqref{eq:refmetric} satisfies all the boundary conditions without introducing excessively large gradients into various functions. To construct such an interpolating function, we first exploit the fact that the $(x,y)$ coordinates \eqref{eq:XYcoords} can be used to cover our entire computational domain. We can therefore globally define a non-compact ``radial'' coordinate, centered at the AdS boundary $(x,y) = (0,1)$, by
\begin{equation}
\hat{R} \coloneqq \sqrt{\left( \frac{x}{1-x} \right)^2 + \left( \frac{1-y}{y} \right)^2} .
\end{equation}
This can be compactified to obtain $\hat{r} \in [0,1]$ via
\begin{equation}
\hat{r} \coloneqq \frac{1}{1+\hat{R}} .
\end{equation}
Note that we have $y \sim \hat{r}$ as $\hat{r} \to 0$ in the inner patch, and $r \sim \hat{r}$ as $\hat{r} \to 1$ in the outer patch. In terms of $\hat{r}$, we may now define our interpolating function as
\begin{equation}
I(\hat{r}) \coloneqq 1-\hat{\varrho}^4 \left( 6-8\,\hat{\varrho}+3\,\hat{\varrho}^2 \right)^2 \quad {\, }\textrm{where} \quad \hat{\varrho} \coloneqq \sin^2 \mathopen{}\left(\mathclose{} \frac{\pi\,\hat{r}}{2} \mathopen{}\right)\mathclose{}\,.
\label{eq:interp}
\end{equation}
Note that, by defining $I$ in terms of $\hat{\varrho}$ rather than $\hat{r}$, we ensure that all even-order normal derivatives of $I$ vanish at $y = 0$ and $r = 1$. We depict our choice of interpolating function in Fig. \ref{fig:interpolating}.

\subsection{Boundary conditions}
\label{subsec:BCs}

The boundary conditions that we impose on our unknown functions are detailed below. Note that at all boundaries there is an additional requirement that the reference metric also satisfies the same conditions. This has already been taken care of in our construction above.

\paragraph{Near region patch:}
\begin{itemize}
\item Horizon $(y=0)$: regularity of the spacetime metric requires that we impose a Neumann boundary condition $\left. \partial_y{\mathcal F} \right|_{y=0}=0$ on all functions, except for $\mathcal W$ which has to vanish. To ensure that the temperature and angular velocity of the AdS rings are the same as those of the AF ring, we further impose $\mathcal Y = \mathcal T$ and $ \mathcal Z=0$ on this boundary.
\item $S^1$ axis $(y=1)$: regularity of the spacetime metric requires that we impose a Neumann boundary condition $\partial_y{\mathcal F}\big|_{y=1}=0$ on all functions, except for $\mathcal W$ which has to vanish. To avoid conical singularities, we further require $\mathcal Y = \mathcal V$ on this boundary. 
\item $S^2$ axes ($x=0$ and $x=1$): again, regularity of the spacetime metric requires that we impose a Neumann boundary condition $\partial_x{\mathcal F}\big|_{x=0,1}=0$ on all functions, except for $\mathcal W$ which has to vanish. To avoid conical singularities, we further require $\mathcal X = \mathcal U$ on these boundaries.
\item Interpolation boundary (blue dots in Fig. \ref{fig:patches} (left)): the value of each function is determined by first interpolating the values of the functions in the outer patch, then applying the coordinate transformations.
\end{itemize}

\paragraph{Far region patch:} 
\begin{itemize}
\item Spacelike infinity $(r=1)$: our ansatz is manifestly asymptotically AdS, provided that we impose a Dirichlet boundary condition $F=0$ on all functions.
\item $S^3$ axes ($a=0$ and $a=1$): regularity of the spacetime metric requires that we impose a Neumann boundary condition $\partial_a F\big|_{a=0,1}=0$ on all functions, except for $W$ which has to vanish. To avoid conical singularities, we further require $Y=V$ at $a=0$ and $Y = U$ at $a=1$.
\item Interpolation boundary (blue dots in Fig. \ref{fig:patches} (right)): the value of each function is determined by first interpolating the values of the functions in the inner patch, then applying the coordinate transformations.
\end{itemize}

These boundary conditions are compatible with the DeTurck vector, $\xi$, vanishing everywhere on the manifold $\mathcal M$. However, we reiterate that for stationary spacetimes we do not have a result analogous to that in \cite{Figueras:2011va} for the static case, so \emph{a priori} our boundary conditions do not necessarily imply that $\xi=0$. \emph{A posteriori}, we have checked that this is indeed the case to within a numerical tolerance.

\subsection{Technical details}
\label{subsec:code}
For the data presented in this paper, the outer patch consists of $320 \times 320$ grid points which are equispaced in both $0.49 \leq r \leq 1$ and $0 \leq a \leq 1$. The inner patch is constructed by taking a $320 \times 320$ grid of points which are equispaced in both $0 \leq x \leq 1$ and $0 \leq y \ \leq 1$, then removing points corresponding to $r > 0.51$. We use either fourth- or sixth-order centered difference stencils to discretise the Einstein--DeTurck equations. The resulting non-linear algebraic system is solved using Newton line-search method with adaptive step size. We used MUMPS \cite{mumps-paper,mumps-manual} or Intel MKL PARDISO \cite{pardiso-paper,pardiso-manual} to solve the linear system at each Newton step. The code is written on top of the PETSc framework \cite{petsc-paper,petsc-manual}. 

We bootstrap our solution procedure by starting at a high temperature, $\tilde{\kappa} \gg 1$, and a ``friendly'' angular velocity, $\tilde{\Omega}_H \approx 2$, where we simply use the reference metric as the initial guess to seed the Newton solver. The $\tilde{\kappa} \gg 1$ condition ensures that the effects of AdS is small, and so the reference metric, which is built from analytically-known AF solutions, is already almost Einstein in both the near-horizon and near-boundary regions. Meanwhile, the $\tilde{\Omega}_H \approx 2$ condition ensures that the geometry interpolating these two regions are not too highly deformed. These two properties combine to give us the best chance of obtaining convergence from Newton's method. Once we have obtained a solution in this somewhat uninteresting regime, we can use it as the initial guess to seed the Newton solver at less ideal parameters. In this way, we can progressively move away from asymptotic flatness towards the more extreme corners of the parameter space.

%%%%%%%%%%%%%%%%%%%%%%%%%%%%%%%%%%%%%%%%%%%%%%% 
%%%%%%%%%%%%%%%%%%%%%%%%%%%%%%%%%%%%%%%%%%%%%%%  
\section{Calculating physical quantities}
\label{sec:physQ}
%%%%%%%%%%%%%%%%%%%%%%%%%%%%%%%%%%%%%%%%%%%%%%% 
%%%%%%%%%%%%%%%%%%%%%%%%%%%%%%%%%%%%%%%%%%%%%%% 

Having constructed the AdS black rings numerically, we now explain how we calculate various physical quantities from our solutions. In particular, we chose a rather nontrivial process to calculate of the rings' mass in order to make sure that we have obtained an accurate answer.

\subsection{Horizon area}

Our boundary condition manifestly makes the hypersurface $H \coloneqq \left\{ x^a \in \mathcal{M} \mid y \equiv 0 \right\}$ a Killing horizon of ${\partial_t - \Omega_H \, \partial_\psi}$. To obtain the horizon's area, we simply need to integrate the volume form pulled back onto a constant-$t$ slice of $H$, thus
\begin{align}
A_H &= \int_{\left. H \right|_t} \dd S \, \sqrt{ \vphantom{g^i_i} \det \left. g_\textrm{\hspace{.1em}inner} \right|_{t, y=0} } \nonumber \\
&= 4\pi^2\int_{0}^{1} \dd x \, \sqrt{\mathcal{X}_0(x,0) \, \mathcal{R}_0(x,0) \, \mathcal{S}_0(x,0)} \, e^{\frac{1}{2} \left[ \mathcal{X}(x,0) + \mathcal{R}(x,0) + \mathcal{S}(x,0) \right]} \, .
\end{align}
We perform the integration using \emph{Mathematica}'s \texttt{NIntegrate} feature, where the unknown functions are first interpolated using polynomial splines.

\subsection{Angular momentum}

Ref. \cite{Magnon:1985sc} showed that the usual AF Komar integral for angular momentum also gives the correct result in AdS asymptotics. Therefore, we calculate
\begin{equation}
J = \frac{1}{16 \pi} \int_{\Sigma} \star \, \dd \mathopen{}\left(\mathclose{} \partial_\psi \mathopen{}\right)\mathclose{}^\flat \, ,
\end{equation}
where $\Sigma$ is any closed spacelike 3-surface bounding a region containing the horizon. The full expression for the integrand is complicated and unenlightening so we will not reproduce it here. We choose $\Sigma$ to reside completely in the outer patch as the hypersurface of constant $r \equiv r_J$, where $r_J$ can be any number. In practice, we found that our result varies by less than 1\% over the range $0.75 < r_J < 0.95$.

\subsection{Mass}
\label{subsec:mass}

We found that the most reliable means of calculating the mass is through the first law of black hole mechanics. In terms of nondimensional quantities this reads
\begin{equation}
\dd\tilde{M} = \frac{\tilde{\kappa}}{8\pi} \, \dd\tilde{A}_{H} + \tilde{\Omega}_{H} \, \dd\tilde{J} .
\label{eq:firstlaw}
\end{equation}
Since each of our data series are obtained by varying $\tilde{\Omega}_{H}$ while keeping $\tilde{\kappa}$ fixed, we can write the above as an ODE
\begin{equation}
\frac{\dd\tilde{M}}{\dd\tilde{\Omega}_{H}} = \frac{\dd}{\dd\tilde{\Omega}_{H}} \lleft \frac{\tilde{\kappa} \, \tilde{A}_{H}}{8 \pi} + \tilde{\Omega}_{H} \tilde{J} \rright - \tilde{J},
\end{equation}
which can be integrated to give
\begin{equation}
\tilde{M}(\tilde{\Omega}_{H}) = \frac{\tilde{\kappa} \, \tilde{A}_{H}(\tilde{\Omega}_{H})}{8 \pi} + \tilde{\Omega}_{H} \tilde{J}(\tilde{\Omega}_{H}) - \int_{\tilde{\Omega}_0}^{\tilde{\Omega}_{H}} \dd\omega \, \tilde{J}(\omega) + \tilde{c}_1 ,
\label{eq:integrate_firstlaw_1}
\end{equation}
where $\tilde{\Omega}_0$ is some chosen limit of integration, and $\tilde{c}_1$ is some constant which depends only on $\tilde{\kappa}$. Clearly, the formula (\ref{eq:integrate_firstlaw_1}) is not of much use unless we also have a way to fix $\tilde{c}_1$ for each $\tilde{\kappa}$. One way to do this is by integrating (\ref{eq:firstlaw}) over a different data series which is continuously connected to an asymptotically flat solution, where we can then use the analytically known mass to fix the constant. To be more precise, we reintroduce explicit factors of the inverse AdS radius $\lambda \coloneqq \ell^{-1}$ into the first law, thus
\begin{align}
\dd \lleft \lambda^2 \, M \rright
&= \frac{\left( \kappa / \lambda \right)}{8\pi} \, \dd \lleft \lambda^3 \, A_H \rright + \left( \Omega_H / \lambda \right) \dd \lleft \lambda^3 \, J \rright \\
&= \dd \mathopen{}\left[\mathclose{} \lambda^2 \left( \frac{\kappa \, A_H}{8\pi} + \Omega_H J \right) \mathopen{}\right]\mathclose{} + \lambda \left( \frac{\kappa \, A_H}{8\pi} + \Omega_H J \right) .
\end{align}
If we now keep $\kappa$ and $\Omega$ fixed while allowing $\lambda$ to vary, this turns into an ODE which we can integrate with respect to $\lambda$, thus
\begin{equation}
M(\lambda) = \left( \frac{\kappa \, A_H(\lambda)}{8\pi} + \Omega_H J(\lambda) \right) + \frac{1}{\lambda^2} \int_0^\lambda \dd\lambda' \left[ \lambda' \left( \frac{\kappa \, A_H(\lambda')}{8\pi} + \Omega_H J(\lambda') \right) \right] + c_2 .
\label{eq:masscalibration}
\end{equation}
To fix the constant $c_2$, note that as $\lambda \rightarrow 0$ the integral term becomes
\begin{equation}
\frac{1}{\lambda^2} \int_0^{\lambda} \dd\lambda' \, \left[ \lambda' \left( \frac{\kappa \, A_H}{8\pi} + \Omega_H J \right)_{\hspace{-0.25em}\lambda=0} + \mathcal{O}(\lambda'^2) \right] \sim \frac{1}{2} \left( \frac{\kappa \, A_H}{8\pi} + \Omega_H J \right)_{\hspace{-0.25em}\lambda=0} \, .
\end{equation}
The AF mass is therefore given by
\begin{equation}
M|_{\lambda=0} = \frac{3}{2} \left( \frac{\kappa \, A_H}{8\pi} + \Omega_H J \right)_{\hspace{-0.25em}\lambda=0} \, ,
\end{equation}
where we have fixed $c_2 = 0$ by identifying the above formula as nothing but the familiar \emph{Smarr relation}. Our mass calculation can thus be summarised as two-step process:
\begin{enumerate}
\item Produce a \emph{``mass calibration''} series of numerical solutions at fixed $\kappa$ and $\Omega_H$, but over a range of $\lambda$. Calculate $A$ and $J$ for each point in the series, then use (\ref{eq:masscalibration}) to calculate $M$. In this paper we somewhat arbitrarily chose $\tilde\kappa = 0.4$ and $\tilde\Omega_H = 0.6$. Note that, in order to calibrate the mass at larger values of $\lambda$, we had to increase both $\kappa$ and $\Omega_H$. This is not a problem provided that $\Omega_H/\kappa$ remains constant, as we can then divide all three parameters ($\kappa$, $\Omega_H$, $\lambda$) by the same number to restore constancy. Using this technique, we were able to fix the mass at $\tilde{\Omega}_{H}/\tilde{\kappa} = 1.5$ for temperatures down to $\tilde{\kappa} = 0.68$.
\item For our \emph{``main''} datasets at some fixed $\tilde{\kappa}$, we can now use (\ref{eq:integrate_firstlaw_1}) to calculate the mass by setting $\tilde{\Omega}_0 = 1.5\tilde{\kappa}$ and $\tilde{c}_1 = \left( \tilde{M} - \frac{\tilde{\kappa}\,\tilde{A}_{H}}{8\pi} - \tilde{\Omega}_{H} \tilde{J} \right)_{\hspace{-0.1em}\tilde{\Omega}_{H} = \tilde{\Omega}_0}$, using the values obtained in step 1. 
\end{enumerate}

We close this section by recalling that \cite{Chrusciel:2006zs} proved that regular black holes solutions in AdS with a single (positive) angular momentum satisfy a BPS bound:
\begin{equation}
J \leq M\,\ell \,.
\end{equation}
 The rotating spherical AdS black holes \cite{Carter:1968ks,Hawking:1998kw,Gibbons:2004uw} and perturbative black rings \cite{Caldarelli:2008pz} satisfy this bound. Indeed, all of our numerical AdS black ring solutions also satisfy this bound. However, since the black ring becomes singular as $J \to M\,\ell$, the solutions that we have managed to construct never get very close to saturating this limit.

%%%%%%%%%%%%%%%%%%%%%%%%%%%%%%%%%%%%%%%%%%%%%%% 
%%%%%%%%%%%%%%%%%%%%%%%%%%%%%%%%%%%%%%%%%%%%%%%  
\section{Geometry}
\label{sec:geometry}
%%%%%%%%%%%%%%%%%%%%%%%%%%%%%%%%%%%%%%%%%%%%%%% 
%%%%%%%%%%%%%%%%%%%%%%%%%%%%%%%%%%%%%%%%%%%%%%% 
In this section we study the geometry of the spatial cross sections of the horizon of the AdS black rings. Throughout this section we shall refer to the size of the rings relative to the the AdS radius $\ell$, as in \S\ref{sec:intro}. For small and short black rings, either thin or fat, the horizon geometry is similar to that of the AF black ring. Therefore, we shall not study the horizon geometry of those rings any further. In what follows, we will describe the geometry of black rings which are either long ($R_{S^1}>\ell$) or large ($R_{S^2}>\ell$). In addition, as we noted in \S\ref{sec:intro}, we have not found any evidence that long rings which are both thin and large exist; in other words, our results suggest that all long thin rings are small. 

The induced metric on the spatial cross sections of the horizon is
\begin{equation}
\dd s_H^2 = R_\parallel(x)^2\,\dd\psi^2 + R_\perp(x)^2\,\dd\phi^2 + \mathcal{X}_0(x,0)\,e^{\mathcal{X}(x,0)}\,\dd x^2\,,
\end{equation}
where $R_\parallel(x) \coloneqq \sqrt{\mathcal{S}_0(x,0)}\,e^{\frac{1}{2}\mathcal{S}(x,0)}$ and $R_\perp(x) \coloneqq \sqrt{\mathcal{R}_0(x,0)}\,e^{\frac{1}{2}\mathcal{R}(x,0)}$. To characterise the geometry, it is useful to consider the the radii of the rotation circle, $R_{S^1}$, and of the transverse two-sphere, $R_{S^2}$. However, unless the ring is very thin, these are rather ambiguously defined because the $S^2$ can be highly distorted. It is possible to come up with some reasonable characterisations of these radii that can provide some information about the actual geometry of the horizon.  Here we follow \cite{Elvang:2006dd}, and define the inner and outer radii of the horizon $S^1$ as
\begin{equation}
R_{S^1}^\textrm{(inner)} \coloneqq \left. R_\parallel \right|_{x=1}\,,\qquad R_{S^1}^\textrm{(outer)} \coloneqq \left. R_\parallel \right|_{x=0}\,.
\end{equation}
There are various ways to characterise the size and shape of the $S^2$. One option is to define the $S^2$ radius as the radius of its equator, where the $S^2$ is fattest, 
\begin{equation}
R_{S^2}^\textrm{(eq)} \coloneqq \max_{0<x<1} \mathopen{}\left(\mathclose{} R_\perp(x) \mathopen{}\right)\mathclose{} \,.
\end{equation}
Alternatively, we can define the radius of the $S^2$ in terms of the proper length of the $S^2$ meridians,
\begin{equation}
R_{S^2}^\textrm{(mer)} \coloneqq \frac{1}{\pi}\int_0^1\,\dd x\,\sqrt{\mathcal{X}_0(x,0)}\,e^{\frac{1}{2}\mathcal{X}(x,0)}\,.
\end{equation}
These two definitions of $R_{S^2}$ coincide when $S^2$ is perfectly round. For thin rings, these two numbers remain very close, however as the rings become fatter neither of these numbers alone provides an authoritative ``size'' of the $S^2$. We can characterise the distortion in the shape of the $S^2$ by defining the \emph{stretch} $\sigma$ as
\begin{equation}
\sigma \coloneqq \frac{R_{S^2}^\textrm{(mer)}}{R_{S^2}^\textrm{(eq)}} - 1 \,.
\label{eq:stretch}
\end{equation}
A perfectly round $S^2$ would therefore have $\sigma = 0$. For the black rings, the gravitational self-attraction means that the $S^2$ is always \emph{prolate}, and so $\sigma \geq 0$. Finally, there is a third definition of the $S^2$ radius in terms of its area
\begin{equation}
R_{S^2}^\textrm{(area)} \coloneqq \sqrt{\frac{A_{S^2}}{4\pi}}\,,
\end{equation}
where the $S^2$ area is given by
\begin{equation}
A_{S^2} \coloneqq 2\pi \int_{0}^{1} \dd x \, R_\perp(x) \, \sqrt{\mathcal{X}_0(x,0)}\,e^{\frac{1}{2}\mathcal{X}(x,0)} \, .
\end{equation}
Note that these three radii are always related by $R_{S^2}^\textrm{(eq)} \leq R_{S^2}^\textrm{(area)} \leq R_{S^2}^\textrm{(mer)}$.

\subsection{Isometric embeddings}
A useful way is to visualise the distortion of the horizon's transverse $S^2$ is to isometrically embed it into 3D Euclidean space $\mathbb E_3$, as was done in \cite{Elvang:2006dd} (see also \cite{Emparan:2014pra}). The metric on the horizon $S^2$ is given by
\begin{equation}
\dd s^2_{S^2} = \mathcal{X}_0(x,0)\,e^{\mathcal{X}(x,0)}\,\dd x^2 + R_\perp(x)^2\,\dd\phi^2\,.
\label{eq:horS2}
\end{equation}
We wish to embed it into $\mathbb E_3$ via cylindrical polar coordinates,
\begin{equation}
\dd s^2_{\mathbb{E}_3}=\dd u^2+\dd\rho^2+\rho^2\,\dd\phi^2\,.
\end{equation}
Letting
\begin{equation}
u=u(x)\,,\qquad \rho = R_\perp(x)\,,
\end{equation}
the resulting induced geometry is given by
\begin{equation}
\dd s^2_\textrm{emb} = \left( R_\perp'(x)^2+u'(x)^2 \right) \dd x^2+R_\perp(x)^2\,\dd\phi^2\,.
\label{eq:embed}
\end{equation}
By comparing \eqref{eq:embed} and \eqref{eq:horS2}, one obtains the embedding
\begin{equation}
u(x) = \int_0^x \dd\eta \sqrt{\mathcal{X}_0(\eta,0)\,e^{\mathcal{X}(\eta,0)} - R_\perp'(\eta)^2}\,,
\end{equation}
which exists for as long as $R_\perp'(\eta)^2 \leq \mathcal{X}_0(\eta,0)\,e^{\mathcal{X}(\eta,0)}$. We find that this condition is satisfied for all the AdS black rings that we have managed to construct. 

\begin{figure}[]
\begin{center}
\begin{subfigure}[c]{0.35\linewidth}
    \centering
    \includegraphics[width=5.5cm]{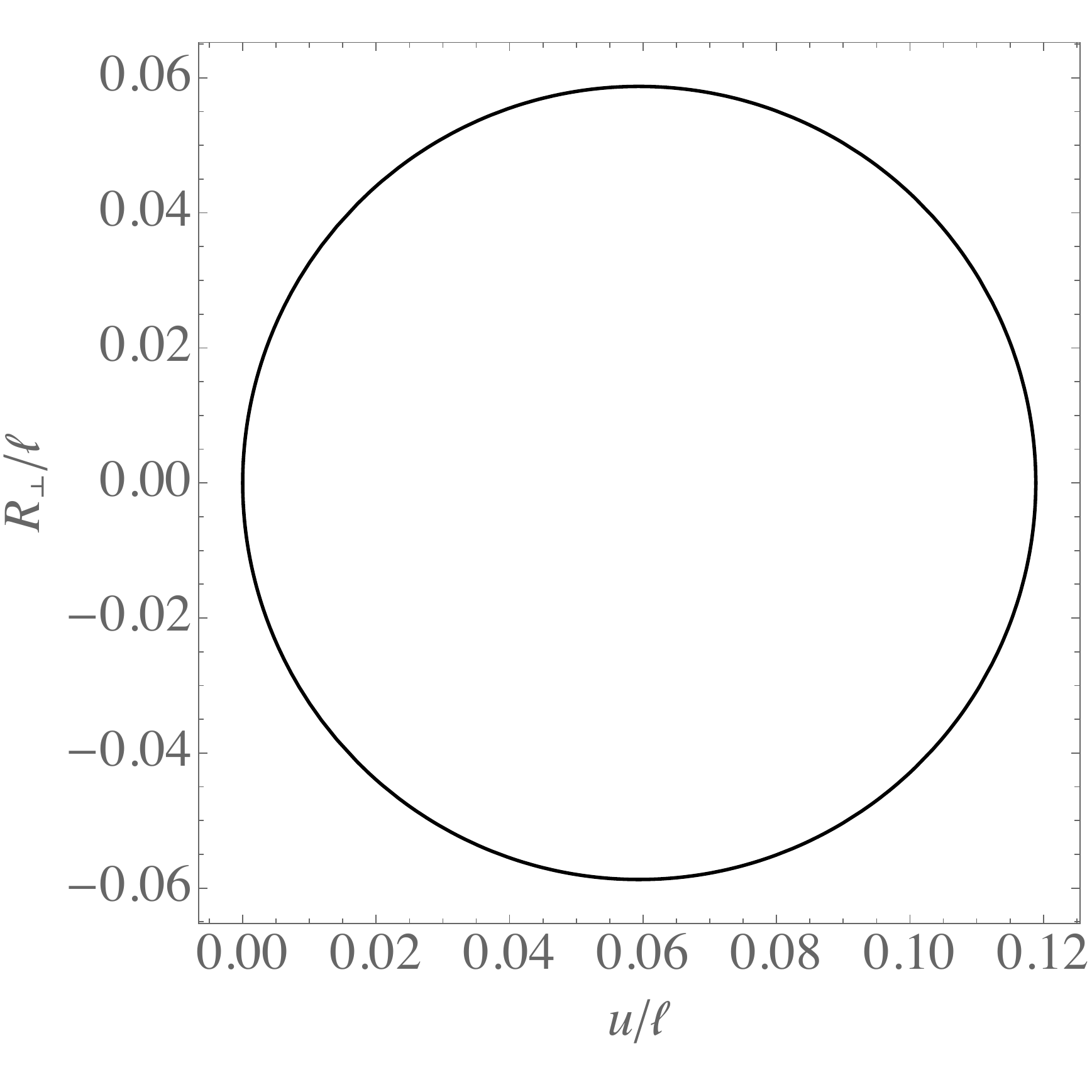}
    \caption{$\kappa\,\ell=5$, $\Omega_H\,\ell=1.3113$}
    \label{subfig:embedThin}
\end{subfigure}%
\begin{subfigure}[c]{0.65\linewidth}
    \centering
    \includegraphics[width=9.5cm]{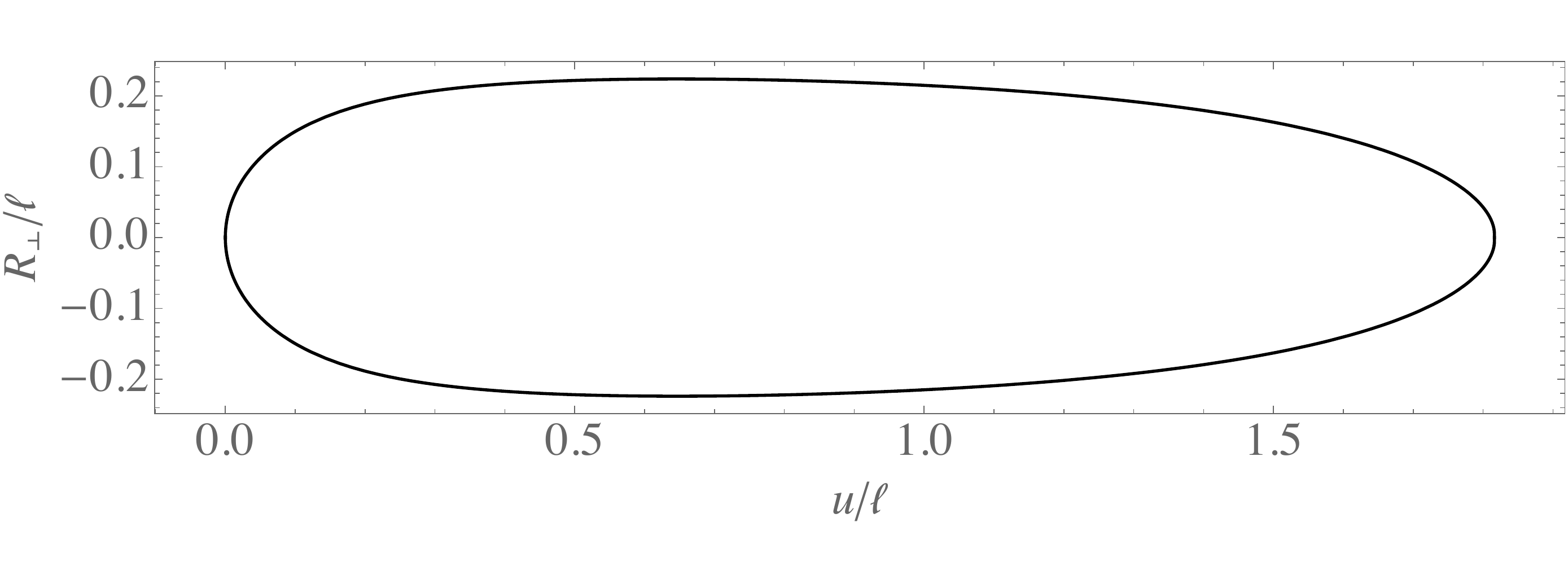}
    \caption{$\kappa\,\ell=0.5$, $\Omega_H\,\ell=1.0375$}
    \label{subfig:embedFat}
\end{subfigure}
\end{center}
\caption{Isometric embeddings. \subref{subfig:embedThin} Embedding of a thin ring with $\tilde{R}_{S^1}^{(\textrm{inner})} = 1.573$ and $\tilde{R}_{S^1}^{(\textrm{outer})} = 2.097$. The stretch \eqref{eq:stretch} is given by $\sigma =0.006$ and hence the horizon $S^2$ is almost perfectly round. \subref{subfig:embedFat} Embedding of a fat ring with $\tilde{R}_{S^1}^{(\textrm{inner})} = 0.285$ and $\tilde{R}_{S^1}^{(\textrm{outer})}= 53.040$. The stretch is $\sigma=1.812$ and hence the horizon $S^2$ is highly deformed from spherical symmetry.}
\label{fig:embedS2}
\end{figure}

In Fig. \ref{fig:embedS2} we present embedding plots of some representative AdS black rings. For long thin rings one would expect that the gravitational self-interaction is small, and hence the horizon $S^2$ should be nearly round. This is precisely what figure \ref{subfig:embedThin} shows for a ring with $\tilde\kappa = 5$ and $\tilde\Omega_H=1.3113$. Note that the gravitational pull due to the negative cosmological constant is compensated by having a large enough angular momentum, and hence it should not affect the geometry of the horizon in a significant manner.  

Fat rings in AdS have a more interesting geometry. Whilst $R_{S^1}^\textrm{(inner)}$ may be small, $R_{S^1}^\textrm{(outer)}$ can be very large, and so in this sense it can be \emph{long}. In Fig. \ref{subfig:embedFat} we depict the $S^2$ embedding of a ring which fits this description: at $\tilde{\kappa}=0.5$ and $\tilde{\Omega}_{H}=1.0375$, we have a rather long $\tilde{R}^\textrm{(outer)}_{S^1}=53.040$. On the other hand, this ring is \emph{not} large in the sense that the typical size of the $S^2$ is not larger than the radius of AdS. For this particular example we have  $\tilde{R}_{S^2}^\textrm{(eq)}=0.224$, $\tilde{R}_{S^2}^\textrm{(area)}=0.415$ and $\tilde{R}_{S^2}^\textrm{(mer)}=0.629$, so the $S^2$ is indeed highly distorted, as Fig. \ref{subfig:embedFat} shows. Note that for any measure of the size of the $S^2$, we have that $\tilde{R}_{S^1}^\textrm{(outer)} \gg \tilde{R}_{S^2}$, so this ring actually looks like a very large and thin membrane with a tiny hole drilled through the middle. It seems reasonable to expect that, by lowering the temperature even further, one should be able to obtain long and fat rings which are also large. But at least in our set up, these are hard to construct numerically. 

\begin{figure}[t]
\begin{center}
\includegraphics[height=10cm]{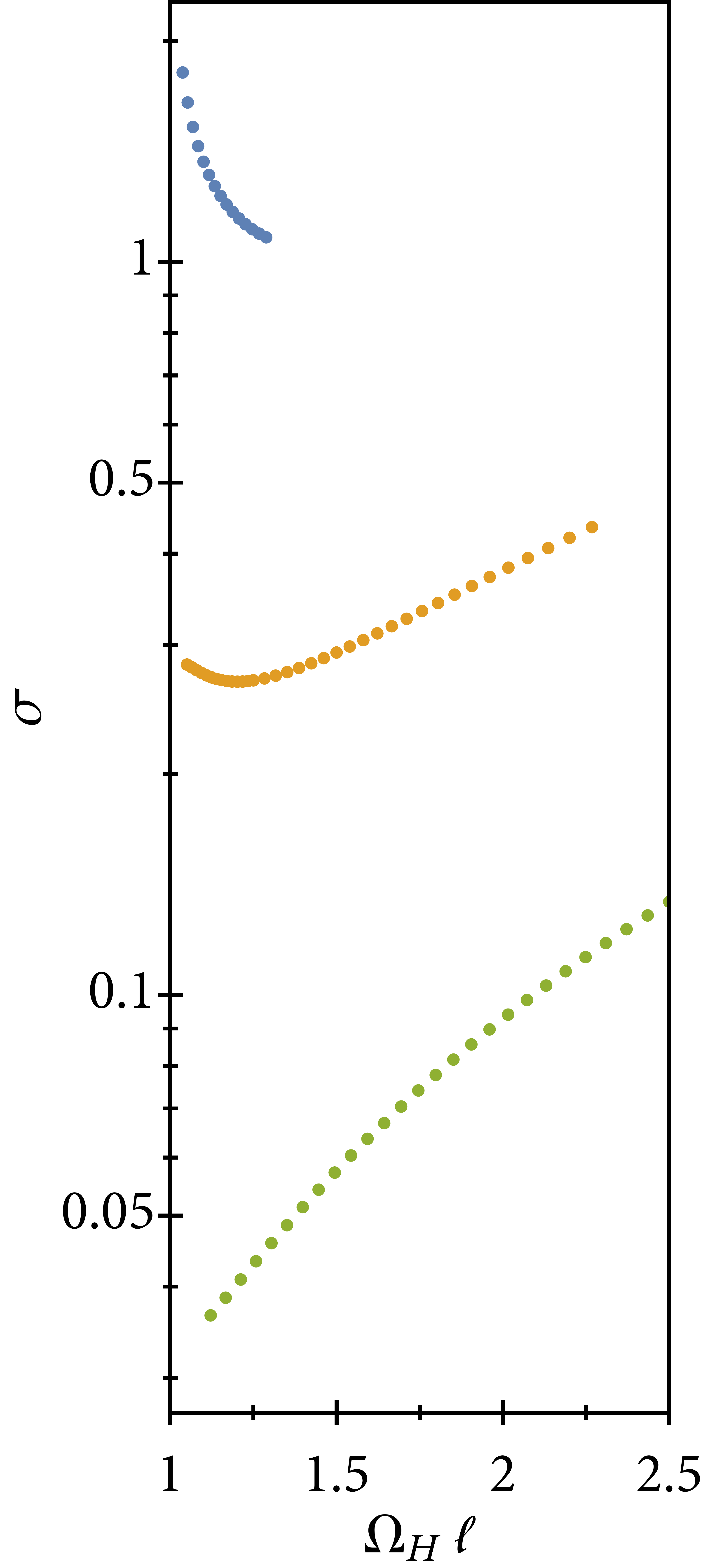}
\end{center}
\caption{The $S^2$ stretch, $\sigma$, plotted against $\tilde\Omega_H$, for rings with $\tilde\kappa=0.5,1,2$ (top to bottom). At high temperatures, the $\tilde\Omega_H\to 1$ limit is reached by thin rings and hence $\sigma\to 0$ in this limit. $\sigma$ increases monotonically as the ring becomes fatter. At low temperatures,  the $\tilde\Omega_H\to 1$ limit is reached by membrane rings, so $\sigma$ cannot be a monotonic function of $\tilde\Omega_H$.}
\label{fig:stretch}
\end{figure}

In Fig. \ref{fig:stretch} we depict the stretch $\sigma$, as defined in \eqref{eq:stretch}, as a function of the angular velocity $\tilde \Omega_H$ for rings at temperatures $\tilde\kappa = 0.5, 1., 2.$ (from top to bottom). At sufficiently high temperatures, the $\tilde\Omega_H\to 1$ limit is reached by thin rings and hence $\sigma\to 0$ in this limit as the $S^2$ becomes perfectly round. As the ring becomes fatter, increasing $\tilde\Omega_H$ while keeping $\tilde\kappa$ fixed, the stretching increases monotonically since the deformation of the $S^2$ also increases. It seems natural to expect that $\sigma$ will diverge in the $\tilde\Omega_H\to \infty$ limit. On the other hand, at sufficiently low temperatures, the $\tilde\Omega_H\to 1$ limit is reached by the membrane rings. For these temperatures, increasing $\tilde\Omega_H$ makes the hole in the middle grow, which implies that $\sigma$ will \textit{decrease} for a while. However, at some point, the ring starts to become fatter again and hence $\sigma$ increases.

\subsection{Invariant radii}

Since $R_\perp(x)$ and $R_\parallel(x)$ are both geometric invariants, plotting them against each other allow us to directly compare the relative sizes of the two cycles. However, the information about the lengths along the $S^2$ meridian is lost and it is therefore important to keep in mind that distances along the curve in these diagrams do not have any real meaning.

Before we describe the geometry of AdS black rings, let us recall some facts about the geometry of rotating AdS black holes. In AdS, the rotating spherical black holes have two different singular limits \cite{Caldarelli:2008pz}. In 5D and for fixed mass, the angular momentum of the spherical black hole is strictly less than the BPS value, $J_\textrm{max}<M\,\ell$. In the limit $J\to J_\textrm{max}$ for fixed $M$, the size of the black hole on the plane of rotation remains finite but the total horizon area goes to zero, hence becoming singular. One can see that in this limit the angular velocity of the horizon diverges. In $D\geq 6$ this corresponds to the well-known ultraspinning limit of black holes, and the value of the angular momentum approaches the BPS value. In AdS, it is possible to take another limit \cite{Caldarelli:2008pz}, even in 5D, in which both the mass $M$ and the angular momentum $J$ diverge whilst their ratio remains finite with $J/(M\,\ell)\to 1$. In this limit, the black hole approaches a rotating black hyperboloid membrane with a horizon topology $\mathbb H^2\times S^1$.

\begin{figure}[t]
\begin{center}
\begin{subfigure}[lt]{0.5\linewidth}
    \centering
    \includegraphics[width=7.5cm]{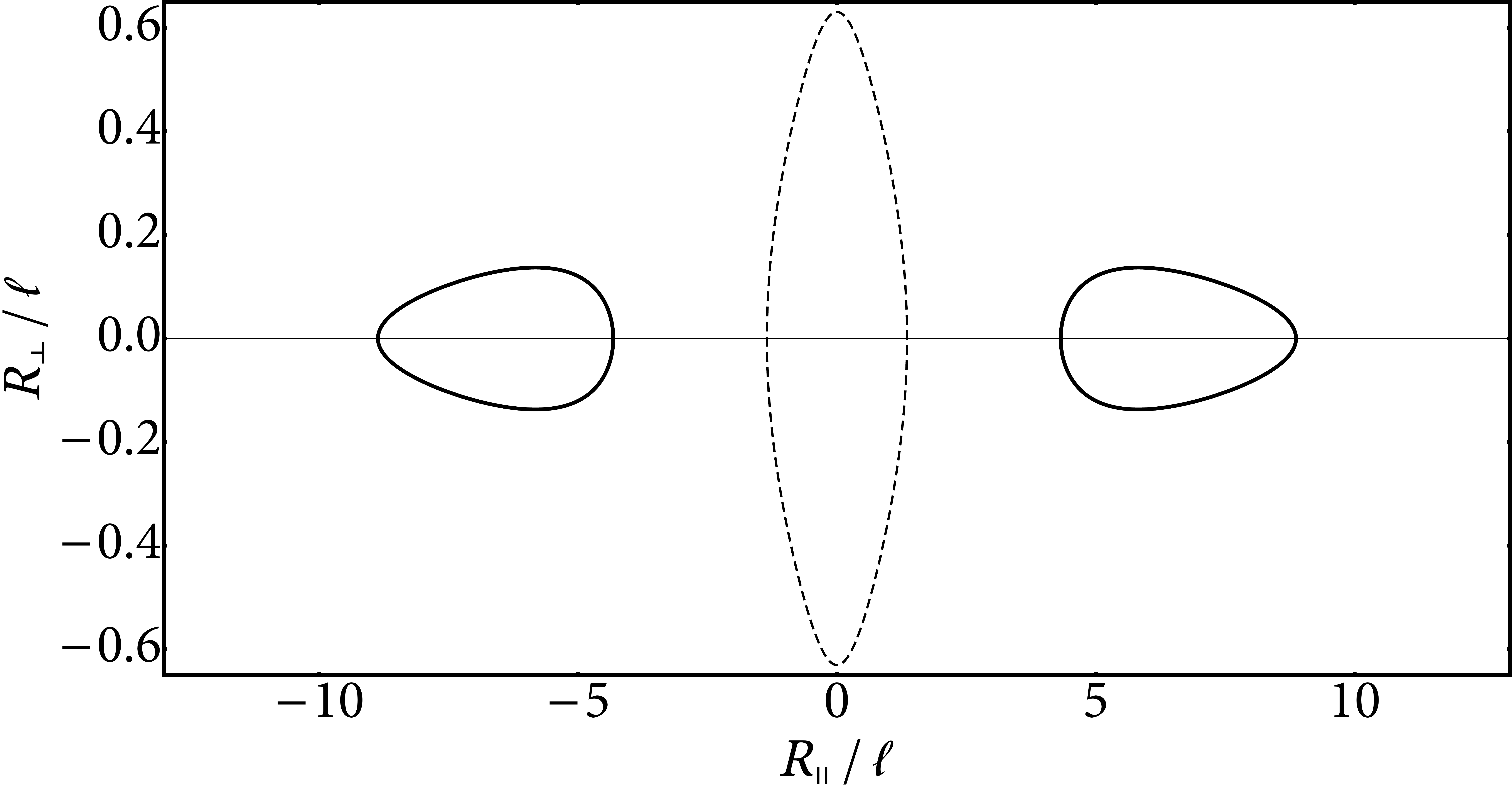}
    \caption{$\kappa\,\ell=2$, $\Omega_H\,\ell=1.07704$}
    \label{subfig:rad1}
\end{subfigure}%
\begin{subfigure}[lt]{0.5\linewidth}
    \centering
    \includegraphics[width=7.5cm]{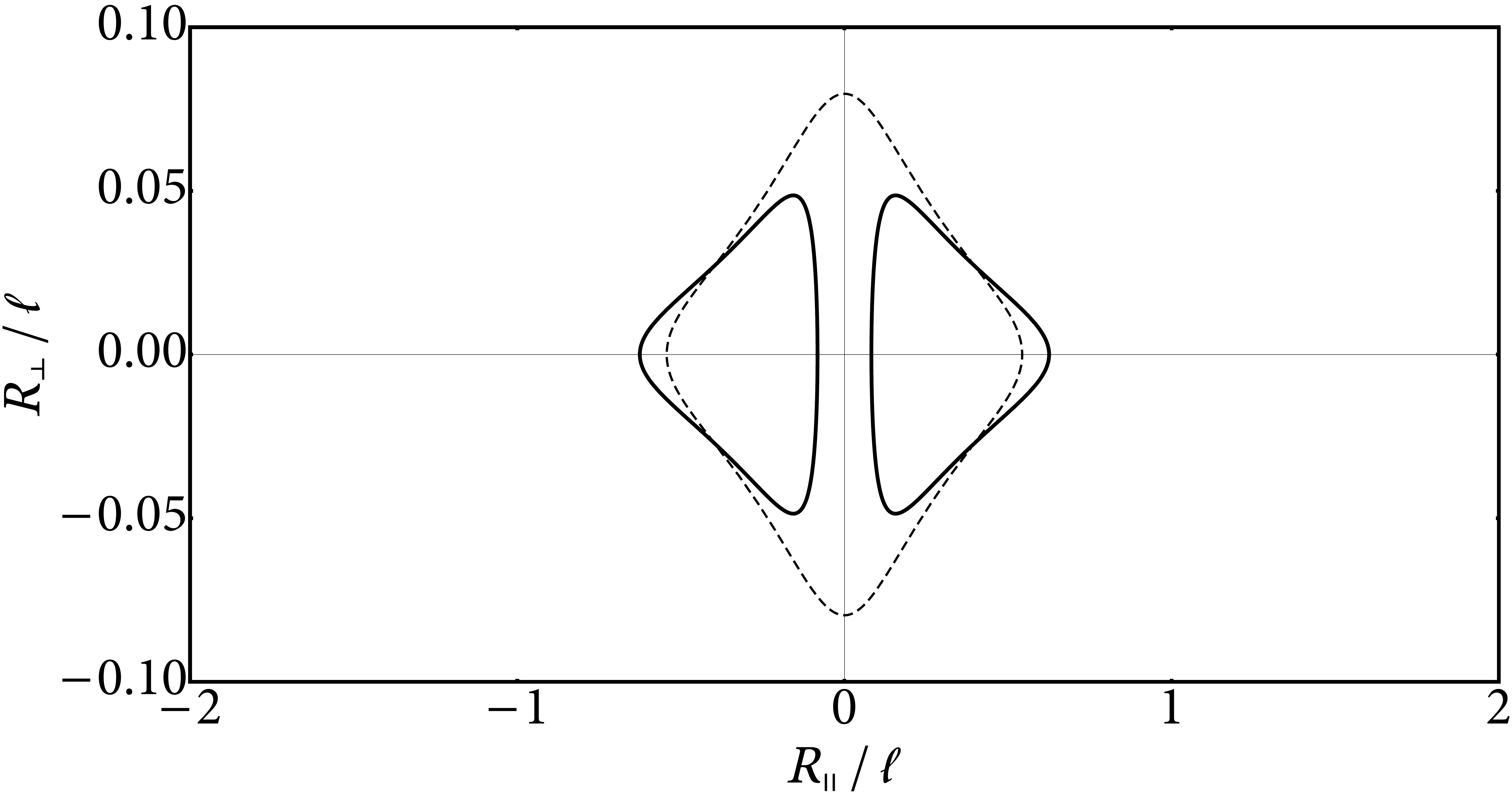}
    \caption{$\kappa\,\ell=2$, $\Omega_H\,\ell=4.53557$}
    \label{subfig:rad2}
\end{subfigure}
\begin{subfigure}[rb]{0.516892\linewidth}
    \centering
    \includegraphics[width=7.65cm]{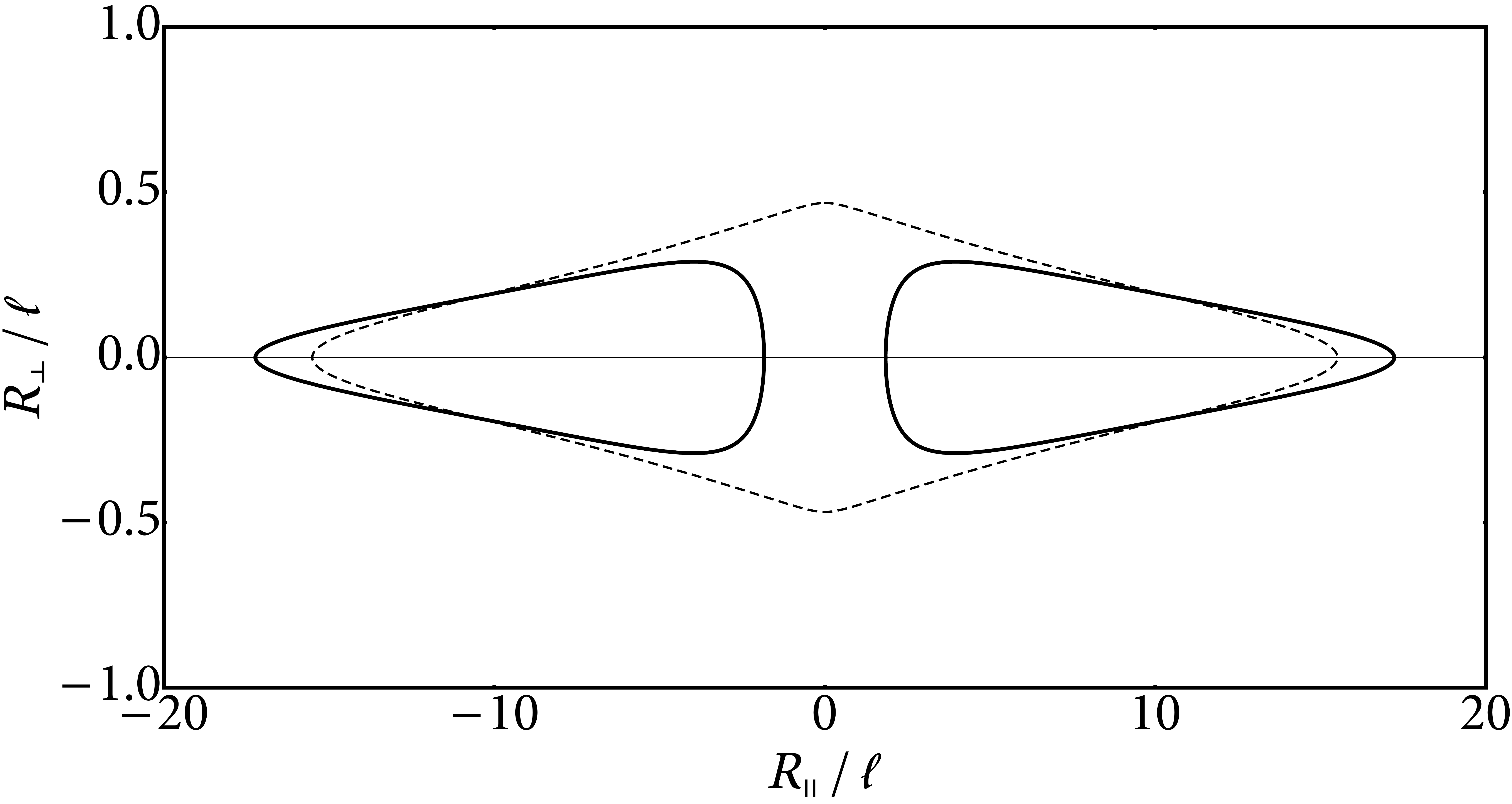}
    \caption{$\kappa\,\ell=1$, $\Omega_H\,\ell=1.05048$}
    \label{subfig:rad3}
\end{subfigure}%
\begin{subfigure}[rb]{0.483108\linewidth}
    \centering
    \includegraphics[width=7.15cm]{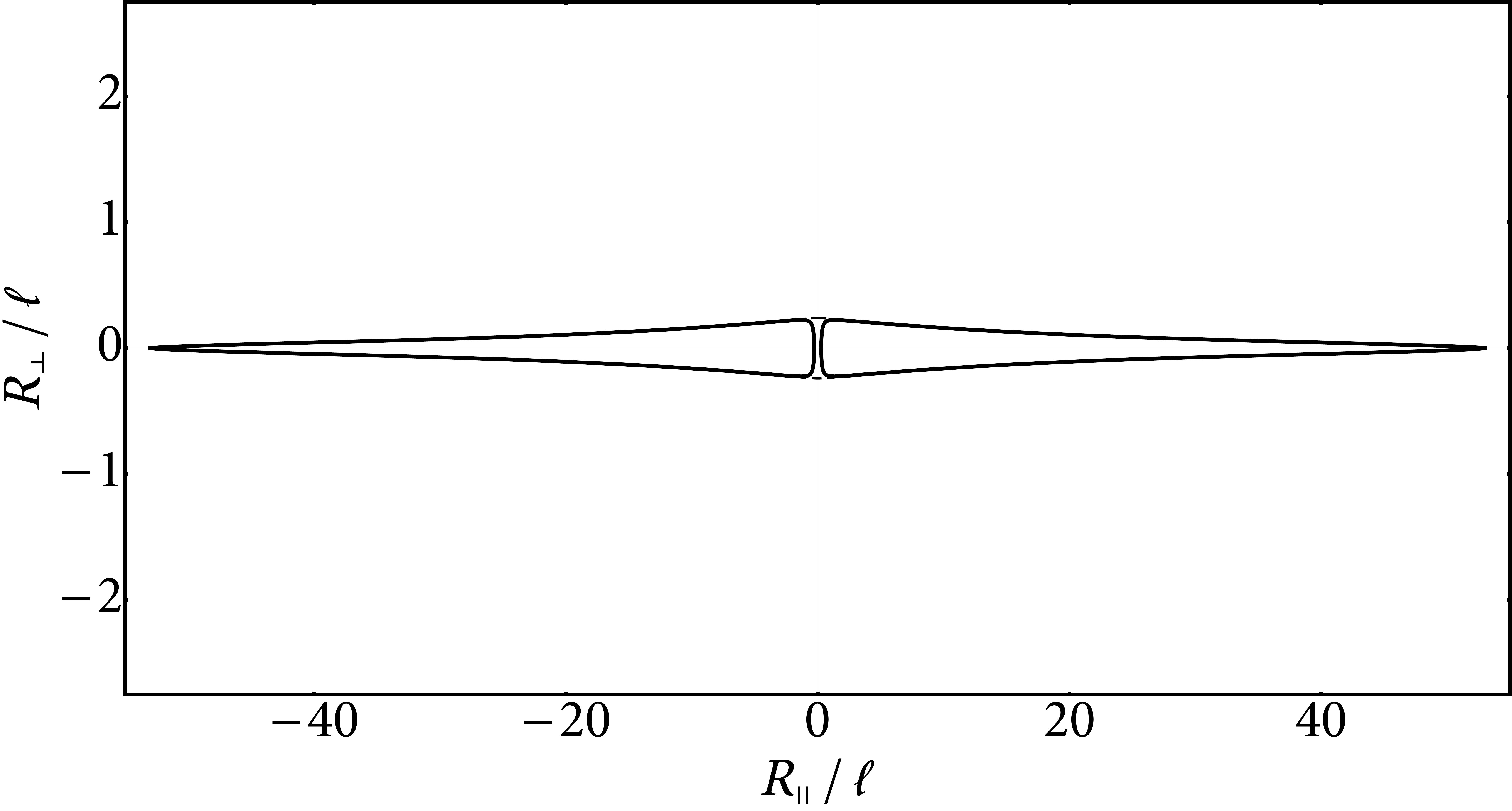}
    \caption{$\kappa\,\ell=0.5$, $\Omega_H\,\ell=1.03750$}
    \label{subfig:rad4}
\end{subfigure}
\end{center}
\caption{Invariant radii plots for some representative temperatures and angular velocities. The solid line corresponds to the black ring and the black dashed line corresponds to the rotating AdS black hole with the same temperature and angular velocity. In \subref{subfig:rad4} we depict the invariant radii for the same black ring as in Fig. \ref{subfig:embedFat}. The aspect ratio is the same in all the plots above, with the vertical axis stretched at exactly 10 times the scale of the horizon axis. }
\label{fig:Rad}
\end{figure}

In Fig. \ref{fig:Rad} we depict some representative plots for rings in different regimes: thin, fat and membrane. We compare the geometry of the ring with that of the rotating AdS black hole with  the same temperature and angular velocity. Note that because we have not fixed the total mass, the actual ``sizes" of the black ring and the black hole can be quite different in certain limits.   Long thin rings are depicted in \ref{subfig:rad1}. As this plot shows, the radius of the $S^1$ of the ring is quite large compared to the radius of AdS, and in some sense the black ring is close to the boundary. As we shall see in \S\ref{sec:stresstensor}, this gets imprinted into the stress-energy tensor of the dual CFT.   In Fig. \ref{subfig:rad2} we show a fat ring; even though we could not reliably construct fatter rings at this particular temperature, this plot suggests that black ring and the black hole are going to merge in the $\tilde\Omega_H\to\infty$ limit. At low enough temperatures, as $\tilde\Omega_H\to 1$ the black ring should tend to the same rotating hyperbolic membrane as does the spherical black hole. This is shown in Figs. \ref{subfig:rad3} and \ref{subfig:rad4}. In particular, in Fig. \ref{subfig:rad4}  it is quite apparent that the black ring and the black hole are tending to the same solution. Note that, since both the ring and the black hole are close to the \textit{same} black membrane, fixing the mass and the angular momentum instead does not produce a significantly different plot. 

%%%%%%%%%%%%%%%%%%%%%%%%%%%%%%%%%%%%%%%%%%%%%%% 
%%%%%%%%%%%%%%%%%%%%%%%%%%%%%%%%%%%%%%%%%%%%%%%  
\section{Thermodynamics of AdS black holes}
\label{sec:thermo}
%%%%%%%%%%%%%%%%%%%%%%%%%%%%%%%%%%%%%%%%%%%%%%% 
%%%%%%%%%%%%%%%%%%%%%%%%%%%%%%%%%%%%%%%%%%%%%%% 

We now move on to discuss the thermodynamics of singly-spinning black holes and black rings in AdS. In \S\ref{sec:canonical} we work in the grand canonical ensemble and study the black hole phases at a fixed temperature and angular velocity. Most of our discussion will lie in this section, as this ensemble is much easier for us to access numerically. We also briefly consider the microcanonical ensemble in \S\ref{sec:microcanonical}, where we instead fix the total mass of the solution. This will allow us to make a direct comparison with the perturbative results of \cite{Caldarelli:2008pz}.

%%%%%%%%%%%%%%%%%%%%%%%%%%%%%%%%%%%%%%%%%%%%%%% 
%%%%%%%%%%%%%%%%%%%%%%%%%%%%%%%%%%%%%%%%%%%%%%%  
\subsection{The grand canonical ensemble}
\label{sec:canonical}
 %%%%%%%%%%%%%%%%%%%%%%%%%%%%%%%%%%%%%%%%%%%%%%% 
%%%%%%%%%%%%%%%%%%%%%%%%%%%%%%%%%%%%%%%%%%%%%%%  

The grand canonical ensemble naturally arises from our numerical procedure, as it is precisely the surface gravity and horizon angular velocity that we are able to fix directly as boundary conditions on the horizon. We first review the properties of the rotating spherical AdS black holes in  \S\ref{subsubsec:canonicalSph}, before looking at the thermodynamics of our numerical black ring solutions in \S\ref{subsubsec:canonicalRing}
.

\subsubsection{Spherical black holes}
\label{subsubsec:canonicalSph}

We begin by looking at 5D asymptotically AdS solutions for which the metric is known analytically. The most trivial of these is of course the pure global AdS solution itself. This is a solution without a horizon. As a result, we can assign an arbitrary period to the Euclidean time coordinate, and so the temperature can be taken to be anything whatsoever. In this context, it is usually referred to as \emph{thermal AdS}. The other class of solutions that we will consider in this subsection are the topologically spherical black holes rotating in a single plane. These are described analytically by the following metric \cite{Hawking:1998kw,Gibbons:2004uw}
\begin{equation}
\begin{gathered}
\begin{aligned}
\dd s^2 &= 
- \frac{\Delta}{\rho^2} \left[ \dd t - \frac{a \sin^2 \theta}{1 - a^2 / \ell^2} \, \dd \psi \right]^2
+ \frac{\Sigma}{\rho^2} \, \sin^2 \theta \left[ a \, \dd t - \frac{r^2 + a^2}{1 - a^2 / \ell^2} \, \dd \psi \right]^2\\
&\phantom{=} + \frac{\rho^2}{\Delta} \, \dd r^2
+ \frac{\rho^2}{\Sigma} \, \dd\theta^2
+ \frac{r^2}{\rho^2} \cos^2 \theta \left[ r^2 \, \Sigma + a^2 \cos^2 \theta \left(1 + \frac{r^2}{\ell^2} \right) \right] \dd\phi^2
\end{aligned} \\
\rho^2 \coloneqq r^2 + a^2 \cos^2 \theta \quad , \quad
\Delta \coloneqq \left( r^2 + a^2 \right) \left( 1 + \frac{r^2}{\ell^2} \right) - 2 \mu \quad , \quad
\Sigma \coloneqq 1 - \frac{a^2}{\ell^2} \, \cos^2 \theta ,
\label{eq:myersperryads}
\end{gathered}
\end{equation}
where $\mu > 0$ is the mass parameter and $a$ is the rotation parameter. Cosmic censorship requires that $| \tilde{a} | < 1$. The event horizon occurs at $r = r_\mathrm{H}$, where $r_\mathrm{H}$ is the largest real root of the polynomial $\Delta(r)$. After nondimensionalising the parameters,
\begin{equation}
\tilde{\mu} \coloneqq \mu / \ell^2 \quad , \quad \tilde{a} \coloneqq a / \ell \quad , \quad \rH \coloneqq r_{\hspace{-0.1em}H} / \ell ,
\end{equation}
the physical quantities for these black holes are given by \cite{Gibbons:2004ai}
\begin{align}
\tilde{M} &= \frac{\pi \, \tilde{\mu} \left( 3 - \tilde{a}^2 \right)}{4 \left( 1 - \tilde{a}^2 \right)^2} , \\%\tag{mass} \\
\tilde{\kappa} &=  \rH \left( 1 + \frac{1 + \rH^2}{\rH^2 + \tilde{a}^2} \right) , \\%\tag{surface gravity} \\
\tilde{A}_{H} &= \frac{2 \, \pi^2 \, \rH \left( \rH^2 + \tilde{a}^2 \right)}{1 - \tilde{a}^2 } , \\%\tag{horizon area} \\
\tilde{\Omega}_{H} &= \frac{\tilde{a} \left( 1 + \rH^2 \right)}{\rH^2 + \tilde{a}^2} , \\%\tag{angular velocity} \\
\tilde{J} &= \frac{2 \, \tilde{a} \, \tilde{M}}{3 - \tilde{a}^2 } . %\tag{angular momentum} 
\end{align}
The solution saturates the BPS bound $| \tilde{J} | \leq \tilde{M}$ as $| \tilde{a} | \rightarrow 1$ (although strictly speaking this is a singular limit). It is easy to verify that these quantities do indeed satisfy the first law of black hole mechanics (\ref{eq:firstlaw}). From now on we will always take $\tilde{a}$ (and hence $\tilde{\Omega}_{H}$ and $\tilde{J}$) to be positive.

We begin by solving for $\tilde{a}$ in terms of $\tilde{\kappa}$ and $\rH$:
\begin{equation}
\tilde{a} = \sqrt{\frac{\rH \left( 1 - \tilde{\kappa} \, \rH + 2 \, \rH^2 \right)}{\tilde{\kappa} - \rH}} .
\label{eq:myersperry-a}
\end{equation}
The BPS limit $\tilde{a} = 1$ corresponds to $\rH = \rH^{(\mathrm{max})} \coloneqq \tilde{\kappa}/2$. This is the upper bound on $\rH$ which holds at all temperatures. The static limit $\tilde{\Omega}_{H} = 0$, i.e. $\tilde{a} = 0$ \emph{and} $\rH \neq 0$, yields two roots $\rH^{(\pm)} \coloneqq \frac{1}{4} \lleft \tilde{\kappa} \pm \sqrt{\tilde{\kappa}^2 - 8} \rright$. The limit $\rH \rightarrow 0$ corresponds to $\tilde{\Omega}_{H} \rightarrow \infty$.

When $\tilde{\kappa} \geq \sqrt{8}$ both $\rH^{(\pm)}$ are real. In this regime, the solutions split into two families: those with $0 < \rH < \rH^{(-)}$ are the \emph{small} rotating black holes, while those with $\rH^{(+)} < \rH < \rH^{(\mathrm{max})}$ are the \emph{large} rotating black holes. There are \emph{no} solutions with $\rH^{(-)} < \rH < \rH^{(+)}$ and so these two families are not connected. 

It is well known \cite{Hawking:1982dh} that, in the presence of a negative cosmological constant, static black holes cannot exist below the critical \emph{Hawking--Page temparature}. This is reflected in our calculation here as the $\tilde{\Omega}_{H} \rightarrow 0$ limit yields imaginary roots when $\tilde{\kappa} < \kappaHP \coloneqq \sqrt{8} \approx 2.828$. However, rotating black holes \emph{can} still exist at these temperatures as long as they are spinning quickly enough. To see this, we substitute (\ref{eq:myersperry-a}) into the expression for $\tilde{\Omega}_{H}$ to obtain
\begin{equation}
\tilde{\Omega}_{H}^2 = \frac{1}{\rH} \left( \tilde{\kappa} - \rH \right) \left( 1 - \tilde{\kappa} \, \rH + 2 \, \rH^2 \right) ,
\end{equation}
and hence
\begin{equation}
\left. \frac{\partial \tilde{\Omega}_{H}^2}{\partial \rH} \right|_{\tilde{\kappa}} = -\frac{1}{\rH^2} \left( \tilde{\kappa} - 3 \, \tilde{\kappa} \, \rH^2 + 4 \, \rH^3 \right) .
\end{equation}
The behaviour of $\tilde{\Omega}_{H}$ is clearly governed by the cubic factor $C( \rH ) \coloneqq \tilde{\kappa} - 3 \, \tilde{\kappa} \, \rH^2 + 4 \, \rH^3$, which is always monotonic in $0 \leq \rH \leq \rH^{(\mathrm{max})}$. While $C(0) = \tilde{\kappa} > 0$ at all temperatures, $C( \rH^{(\mathrm{max})} ) = \frac{1}{4} \, \tilde{\kappa} \left( 4 - \tilde{\kappa}^2 \right)$ changes sign at $\tilde{\kappa} = 2$. The low temperature solutions are therefore further split into two regimes.

For $2 < \tilde{\kappa} < \kappaHP$ we have $C( \rH^{(\mathrm{max})} ) < 0$, so $C$ must have a root $\rH^*$ corresponding to the turning point in $\tilde{\Omega}_{H}$. Solutions with $\rH < \rH^*$ are the low-temperature continuation of the \emph{small} black holes family, while those with $\rH > \rH^*$ are the continuation of the \emph{large} black holes family. These two branches \emph{are} now continuously connected to each other at these temperatures.

For $0 < \tilde{\kappa} < 2$ the function $\tilde{\Omega}_{H}$ has no turning point. Instead, it is monotonically decreasing, with $\tilde{\Omega}_{H} \rightarrow 1$ as $\rH \rightarrow \rH^{(\mathrm{max})}$. Physical quantities of these solutions behave like those of the \emph{small} black holes, while the large black holes cease to exist in this temperature regime. Note, however, that these ``small'' black holes can still grow to arbitrarily large horizon areas as $\tilde{\Omega}_{H} \searrow 1$.

\begin{figure}[t]
\begin{center}
\includegraphics[height=7.5cm]{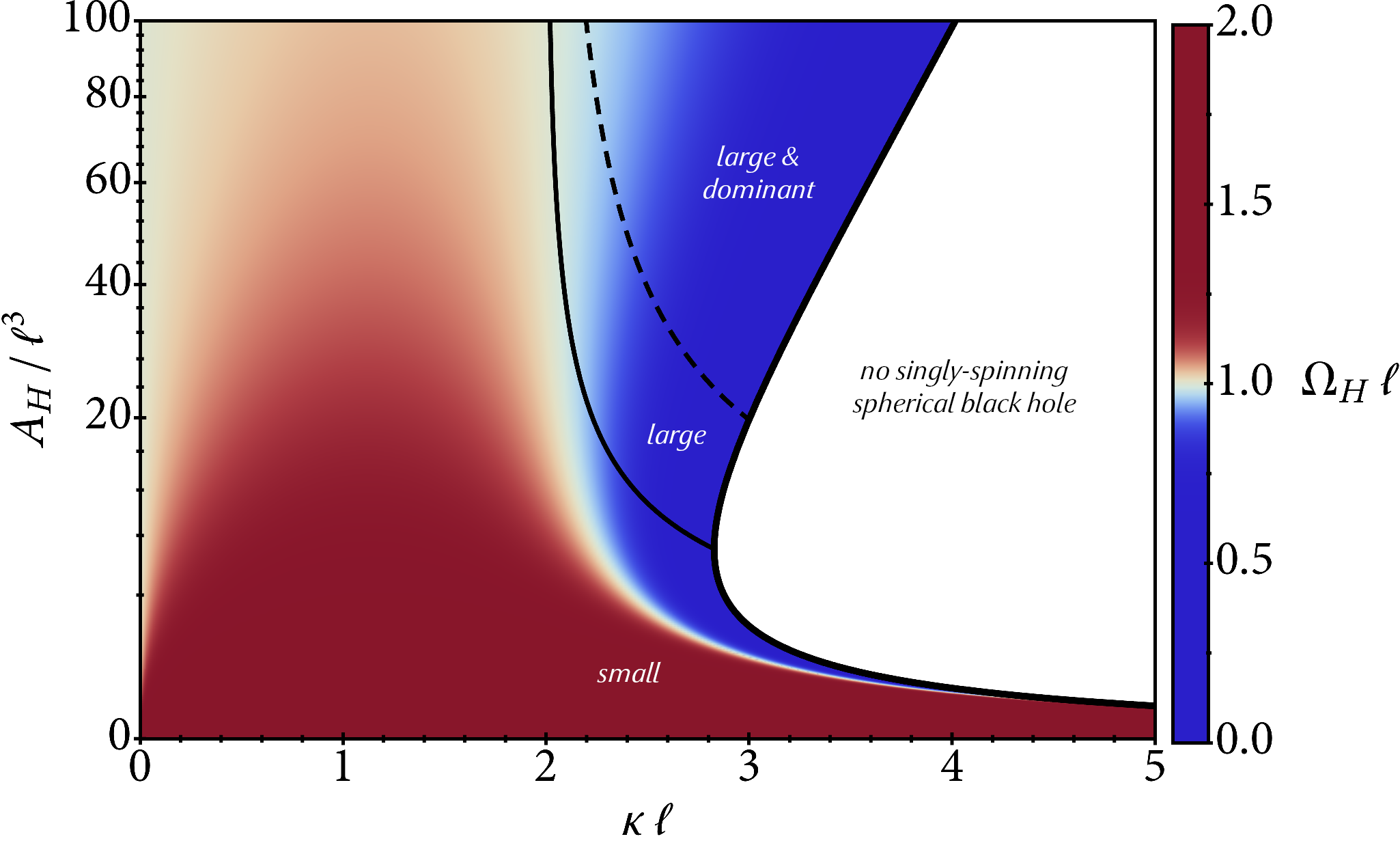}
\end{center}
\caption{A phase diagram for rotating spherical AdS black holes. The colour shows the horizon's angular velocity $\tilde{\Omega}_H$ for each black hole solution, each uniquely parametrised by $( \tilde{\kappa}, \tilde{A}_{H} )$. Blue-tinted points are \emph{superradiant-stable} solutions with $\tilde{\Omega}_H < 1$. The coloured region is bounded on the right by a thick curve corresponding to the static solutions, which only exist when $\tilde{\kappa} \geq \sqrt{8}$. The solid interior curve marks the boundary between the \emph{small} and \emph{large} black holes, the latter of which cease to exist when $\tilde{\kappa} \leq 2$. Instead, in this regime ``small'' black holes can have an arbitrarily large $\tilde{A}_{H}$. Lastly, solutions above the dashed interior curve (all of which are \emph{large} black holes) have negative grand canonical potential $\tilde{\Phi}$, and thus dominate the ensemble. }
\label{fig:AllowedArea}
\end{figure}

We can now discuss the thermodynamics of these solutions. Fixing the values of $\tilde{\kappa}$ and $\tilde{\Omega}_{H}$ is analogous to placing the system in the grand canonical ensemble, and thus phase dominance is determined by the grand canonical potential $\tilde{\Phi} \coloneqq \tilde{M} - \frac{1}{8\pi} \, \tilde{\kappa} \, \tilde{A}_{H} - \tilde{\Omega}_{H}\,\tilde{J}$. With our normalisation, thermal AdS obviously has $\tilde{\Phi} = 0$.  The small black holes always have a positive $\tilde{\Phi}$ and therefore never dominates the ensemble.  On the other hand, large rotating black holes with negative $\tilde{\Phi}$ do exist at \emph{all} $\tilde{\kappa} > 2$. In the range $2 < \tilde{\kappa} < 3$, some large black holes still have a positive $\tilde{\Phi}$ and so there is an angular velocity threshold below which thermal AdS is still dominant. When $\tilde{\kappa} > 3$, even the static solution has $\tilde{\Phi} < 0$ and so the entire large black hole branch becomes dominant. It is important to note that large rotating black holes always obey the Hawking--Reall bound, $\tilde\Omega_H<1$, and hence they should be classically dynamically stable. On the other hand, small black holes with $\tilde\Omega_H>1$ should be unstable under superradiance. The actual phase diagram of rotating AdS black holes is summarised in Fig. \ref{fig:AllowedArea}.

\subsubsection{Black rings}
\label{subsubsec:canonicalRing}

\begin{figure}[t]
\begin{center}
\includegraphics[width=10cm]{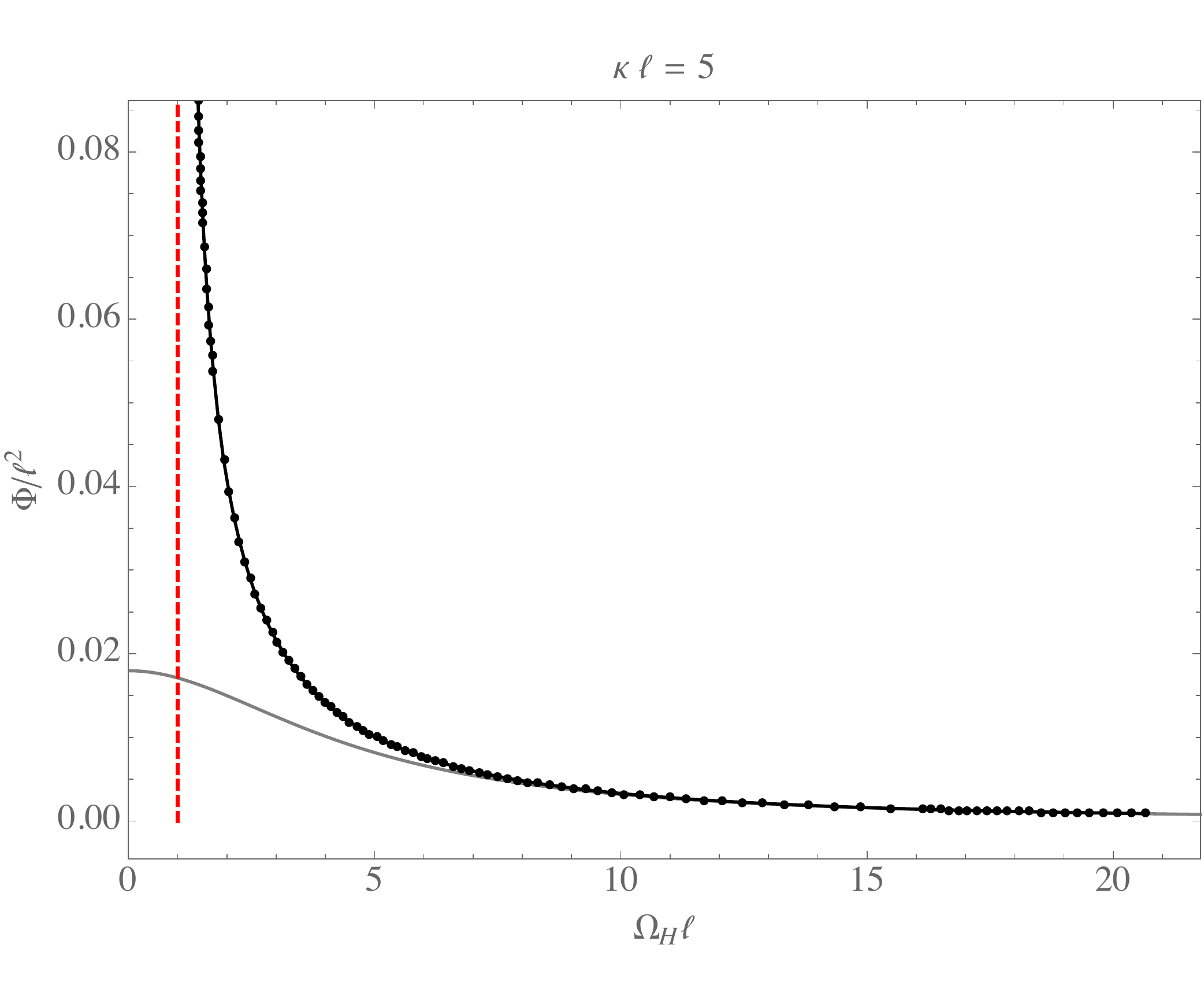}
\end{center}
\caption{Grand canonical potential for AdS black rings (black dots) and small AdS black holes (gray curve) for $\tilde\kappa=5$. The dashed red line indicates the $\tilde\Omega=1$ limit. AdS black ring never dominate the grand canonical ensemble and their angular velocity always exceeds the Hawking--Reall bound.}
\label{fig:grandpotential}
\end{figure}

Using the procedure described in \S\ref{sec:physQ} to accurately calculate the physical quantities for the AdS black rings, it is straightforward to compute the associated grand canonical potential. In Fig. \ref{fig:grandpotential} we depict the grand canonical potential for representative AdS black rings at $\tilde\kappa=5$, as a function of the angular velocity $\tilde\Omega_H$, and compare it with that of the \textit{small} rotating AdS black holes at the same temperature. For these configurations, the grand canonical potential of the large black holes is always negative and off the scale of this plot.  At \textit{any} other non-zero temperature, the picture for the rings is qualitatively the same. The only difference  in the phase diagram for rings as one varies $\tilde\kappa$ is that, for $\tilde\kappa>2$, the $\tilde\Omega\searrow 1$ limit is attained by thin rings, whilst for $\tilde\kappa<2$ this limit is attained by membrane rings. Therefore, we conclude that black rings in AdS, regardless of their size or shape, \textit{never} dominate the grand canonical ensemble. Moreover, in the $\tilde\Omega_H\to\infty$ limit, which is always attained from the fat branch, black rings are connected to the small rotating AdS black holes. Hence, from a thermodynamic point of view, black rings behave in a similar manner as small rotating black holes. In particular, they are always thermodynamically unstable. 

%%%%%%%%%%%%%%%%%%%%%%%%%%%%%%%%%%%%%%%%%%%%%%% 
%%%%%%%%%%%%%%%%%%%%%%%%%%%%%%%%%%%%%%%%%%%%%%%  
\subsection{The microcanonical ensemble}
\label{sec:microcanonical}
%%%%%%%%%%%%%%%%%%%%%%%%%%%%%%%%%%%%%%%%%%%%%%% 
%%%%%%%%%%%%%%%%%%%%%%%%%%%%%%%%%%%%%%%%%%%%%%% 

Accessing the microcanonical ensemble is rather more challenging from the numerical point of view. This is because the mass of the black hole is defined in terms of an integral over some hypersurface in the spacetime, which is not a local condition at any given point. We are not aware of any boundary condition which would allow for a specific mass to be fixed directly. For a given mass, we instead resort to first estimating the combinations of the parameters $(\tilde{\kappa}, \tilde{\Omega}_{H})$ by interpolating the $\tilde{M}$ across the datasets that we produced for the grand canonical ensemble. Having obtained solutions near the desired mass at various values of $\tilde\kappa$, we then proceed to fine-tune our estimation by performing a bisection search on $\tilde{\Omega}_{H}$ until the mass becomes correct to within some tolerance. This entails a significant amount of work for each data point (on average we had to obtain four full solutions for each bisection search), and therefore we only present here a phase diagram for one particular choice of $\tilde{M}$.

We choose to focus on $\tilde{M} = 10$ as this allows us to complete the perturbative picture presented in Figure 1 of \cite{Caldarelli:2008pz}. For this mass, we have obtained ten solutions in the range $0.6 \leq \tilde\kappa \leq 2.0$, and plotted their horizon areas $\tilde{A}_{H}$ against the angular momenta $\tilde{J}$. This can be superposed onto the corresponding curve for the spherical black holes at the same mass, and also the perturbative result of \cite{Caldarelli:2008pz}. As expected, our data points approach the perturbative curve at larger values of $\tilde\kappa$, for which the ring is geometrically thin. The BPS bound is approached as $\tilde\kappa \rightarrow \infty$, where we have $\tilde{A}_{H} \rightarrow 0$, $\tilde{J} \rightarrow \tilde{M}$, and the ring becomes arbitrarily thin. Similarly to the AF black rings, the $(\tilde{A}_{H},\tilde{J})$ curve for the AdS black rings has a cusp separating ``\emph{fat}'' and ``\emph{thin}'' rings in the thermodynamical sense. We estimate that for $\tilde{M} = 10$ this occurs at $\tilde\kappa^* \approx 0.93$. As $\tilde\kappa$ decreases beyond $\tilde\kappa^*$, the ring becomes fatter and the curve approaches that of the spherical black hole, before merging at the singular solution at $\tilde\kappa = 0$.

\begin{figure}[t]
\begin{center}
\includegraphics[width=10cm]{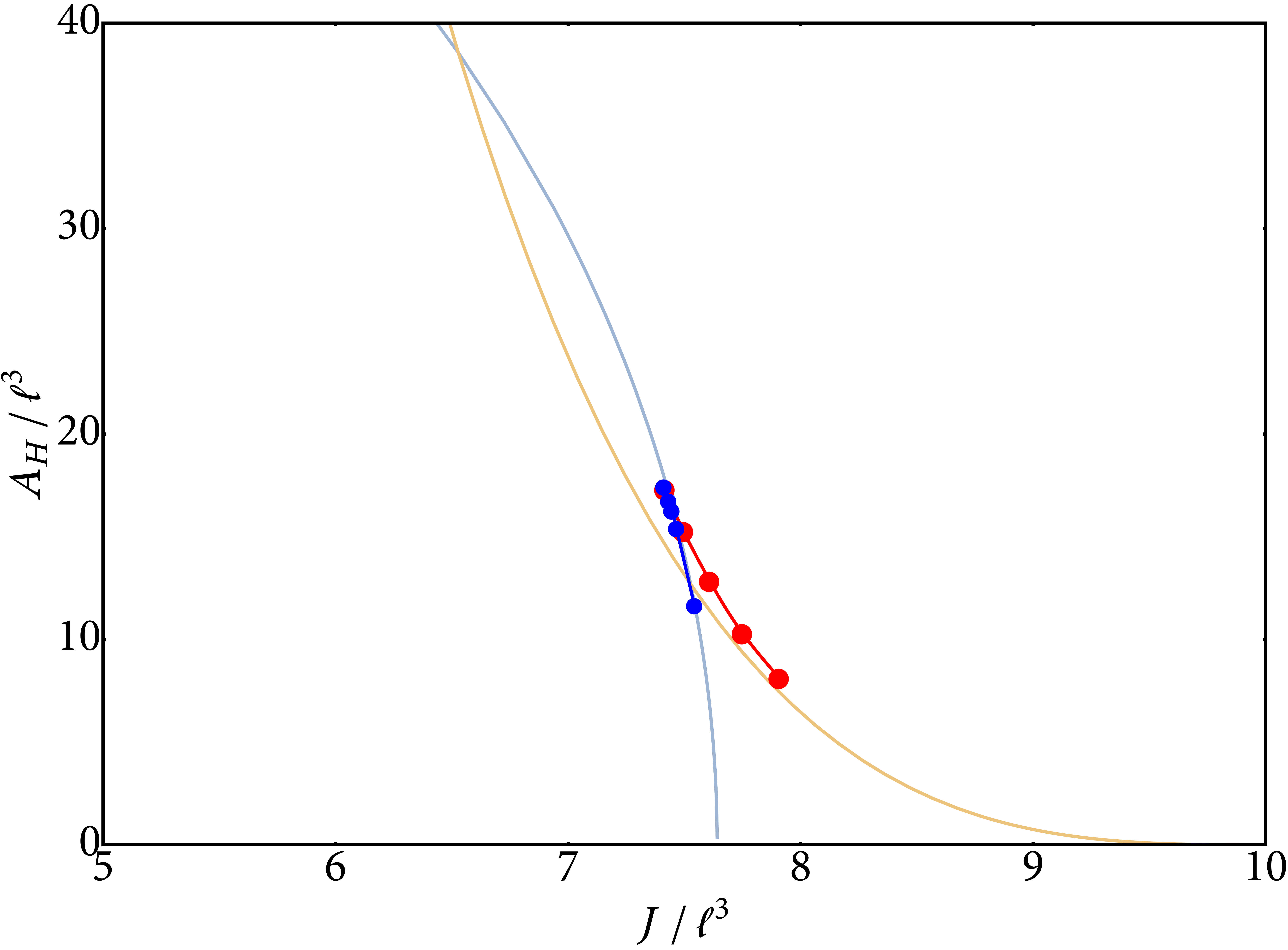}
\end{center}
\caption{Microcanonical ensemble phase diagram for mass $\tilde{M} = 10$, showing the horizon area $\tilde{A}_H$ against angular momentum $\tilde{J}$. The faint blue curve in the background is the exact result for the spherical AdS black holes, while the yellow one shows the perturbative results of \cite{Caldarelli:2008pz}. Red points show our black rings with $\tilde{\kappa} \geq 1$, i.e. the \emph{thin} rings, while the blue points show rings with $\tilde{\kappa} < 1$, i.e. the \emph{fat} rings. There is a cusp at the minimum value of $\tilde{J}$ around $\tilde{\kappa}^* \approx 0.93$. Our numerical results approach the perturbative curve at high $\tilde{\kappa}$, while for low $\tilde{\kappa}$ the fat rings approach the curve for spherical black holes.}
\label{fig:micro}
\end{figure}

The existence of the cusp allows us to precisely separate \emph{thin rings} from \emph{fat rings} in the microcanonical ensemble. We define \emph{fat rings} by requiring
\begin{equation}
\left. \frac{\partial\tilde{A}_{H}}{\partial\tilde{\kappa}} \right|_{\tilde{M}} > 0 \, .
\end{equation}
One can easily apply the first law (\ref{eq:firstlaw}) and the chain rule to deduce that this is equivalent to 
\begin{align}
\Delta \coloneqq \left. \frac{\partial \tilde{A}_{H}}{\partial \tilde{\Omega}} \right|_{\tilde{\kappa}} \left. \frac{\partial \tilde{J}}{\partial \tilde{\kappa}} \right|_{\tilde{\Omega}}
- \left. \frac{\partial \tilde{A}_{H}}{\partial \tilde{\kappa}} \right|_{\tilde{\Omega}} \left. \frac{\partial \tilde{J}}{\partial \tilde{\Omega}} \right|_{\tilde{\kappa}} 
> 0 \, .
\end{align}
It was shown in \cite{Monteiro:2009tc} that the above quantity is proportional to the Hessian determinant of the grand canonical potential $\tilde{\Phi}$ with respect to the coordinates $(\tilde{\kappa},\tilde{\Omega})$. Note also that $\Delta$ carries the same sign as the constant-$J$ heat capacity, $C_J$.

%%%%%%%%%%%%%%%%%%%%%%%%%%%%%%%%%%%%%%%%%%%%%%% 
%%%%%%%%%%%%%%%%%%%%%%%%%%%%%%%%%%%%%%%%%%%%%%%  
\section{Holographic stress tensor}
\label{sec:stresstensor}
%%%%%%%%%%%%%%%%%%%%%%%%%%%%%%%%%%%%%%%%%%%%%%% 
%%%%%%%%%%%%%%%%%%%%%%%%%%%%%%%%%%%%%%%%%%%%%%% 
 In this section we study the stress tensor of ${\mathcal N} = 4$ SYM  on $\mathbb R_t\times S^3$ for states which are dual to rotating AdS black holes and black rings. For rotating spherical black holes in AdS this was first done in \cite{Bhattacharyya:2007vs} and we shall borrow some results from this reference. 
 
 We extract the stress tensor using the standard holographic renormalisation prescription \cite{deHaro:2000xn}. Note that with our choice of outer patch, see \eqref{eq:metricFar}, the boundary geometry is given by the standard metric on the Einstein static universe, $\mathbb R_t\times S^3$. As is well known, ${\mathcal N} = 4$  SYM on this geometry has a non-zero Casimir contribution \cite{Horowitz:1998ha}. In the derivatione below we will have to subtract this universal piece. 
 
In order to extract the stress tensor of the dual CFT, we must first transform the outer region metric $\dd s^2_\mathrm{outer}$ from \eqref{eq:metricFar} into Fefferman--Graham coordinates. In these coordinates, $\dd s^2_\mathrm{outer}$ can be expanded around $z=0$ in the form
\begin{gather}
\dd s^2_\mathrm{outer} \sim \frac{\ell^2}{z^2}\,\left(\dd z^2 + \left[ g_{(0)}(x) + z^2 \, g_{(2)}(x) + z^4 \, g_{(4)}(x) + \mathcal{O} \mathopen{}\left(\mathclose{} z^5 \mathopen{}\right)\mathclose{} \right]_{ij} \dd x^i\,\dd x^j\right)\,, \\
g_{(0)}(x) + z^2 \, g_{(2)}(x) \coloneqq \mathrm{diag} \mathopen{}\left[\mathclose{} -1-{\textstyle \frac{z^2}{2\,\ell^2}}, \, \ell^2-{\textstyle \frac{z^2}{2}}, \, \cos^2\theta \left( \ell^2-{\textstyle \frac{z^2}{2}} \right), \, \sin^2\theta \left( \ell^2-{\textstyle \frac{z^2}{2}} \right)  \mathopen{}\right]\mathclose{} \, ,  \nonumber \\
x^i \coloneqq (t, \theta, \phi, \psi) \, . \nonumber
\label{eq:FGmetric}
\end{gather}
Note that index contractions are with respect to the metric $g_{(0)}$, which in this case is just the standard metric on the Einstein static universe. Then, the v.e.v. of the dual stress tensor is given by \cite{deHaro:2000xn}
\begin{equation}
\langle T_{ij}\rangle = \frac{\ell^3}{4\,\pi\,G_5}\left( g_{(4)ij}-\frac{1}{8} 
\left[\left(\trace g_{(2)}\right)^2 - \trace g_{(2)}^2  \right] g_{(0)ij}
-\frac{1}{2}\,\big( g_{(2)}^2 \big)_{ij}+\frac{1}{4} \left[ \vphantom{g_{(2)}^2} \trace g_{(2)} \right] g_{(2)ij} \right)\,.
\end{equation}
We perform the change into FG coordinates by substituting $r \to r(z,\theta)$ and $a \to a(z,\theta)$ into \eqref{eq:metricFar} and then imposing the Einstein--DeTurck equations near $z=0$ at each order in $z$ up to (and including) $\mathcal{O}(z^5)$. We also subtract off the contribution from pure global AdS, i.e. the Casimir energy. The final result, expressed in terms of our unknown functions, is given by
\begin{equation}
\begin{gathered}
\left\langle T_{ij}\right\rangle^\textrm{sub}\,\dd x^i\,\dd x^j=\frac{N_c^2 \, k^8}{768 \, \pi^2 \, \ell^8}\,
\bigg[ \phantom{} - T^{(4)}(\theta)\,\dd t^2 -2\,\ell\,\Xi(\theta)\,\dd t\,\dd\psi \hspace{4cm} \\
\hspace{4cm} \phantom{} + \ell^2\left(
Y^{(4)}(\theta)\,\dd \theta^2
+\cos^2\theta\,U^{(4)}(\theta)\,\dd\phi^2
+\sin^2\theta\,V^{(4)}(\theta)\,\dd\psi^2
\right) \bigg]\,,
\end{gathered}
\label{eq:stress}
\end{equation}
where $F^{(4)}(\theta) \coloneqq \partial_r^4 F \big(r,\frac{2 \, \theta}{\pi} \big)|_{r=k}$ for any function $F$, and
\begin{equation}
\begin{gathered}
\Xi(\theta) \coloneqq \frac{3 \, \pi^4 \, \Gamma \, \big( \Gamma + \tilde{\Omega}_H \big)^2 \, \big[ \, \big( \Gamma + \tilde{\Omega}_H \big) - \big( \Gamma - 3 \, \tilde{\Omega}_H \big) \, \cos 2\theta \, \big]^2 \, \sin^2 \theta}{8 \, k^4 \, \big( \Gamma - \tilde{\Omega}_H \big)^4} \, \bigg( 1 + Z\big(k,{\textstyle \frac{2\,\theta}{\pi}} \big) \bigg) \, , \\
\textrm{with} \quad \Gamma \coloneqq \sqrt{4\tilde{\kappa}^2 + \tilde{\Omega}^2_H} \, .
 \end{gathered}
 \end{equation}
Note that our differentiation variable $r$ has the dimensions of length, and we have reinstated the compactification scale $k$, as defined in \eqref{eq:compactification}, so that the expression above manifestly has the correct dimensions. The expression has already been somewhat simplified by using the relation $T^{(4)}(\theta) + Y^{(4)}(\theta) + U^{(4)}(\theta) + V^{(4)}(\theta) = 0$ which arises from Einstein's equation at $\mathcal{O}(z^2)$. However, we have \emph{not} completely eliminated any one of the four functions out altogether, so $\left\langle T_{ij} \right\rangle^\mathrm{sub}$ is not manifestly traceless. This will prove useful for our calculation method as detailed in \S\ref{subsec:extraction}. From now on all our stress tensors will have the Casimir contribution subtracted, and thus we no longer explicitly show the ``sub'' superscripts.
 
\subsection{Accurate extraction of the stress tensor}
\label{subsec:extraction}

Having derived the stress tensor components in terms of our unknown functions, we now explain our method of actually evaluating \eqref{eq:stress} numerically. For each unknown function $F$, we first apply the following protocol. At each $a \in [0,1]$ in the outer patch grid, we take eight data points closest to the boundary at $r = 1$ and fit onto them a polynomial of the form $p_a(r) \coloneqq \frac{1}{4!} \, \alpha_a \, (1-r)^4 + \frac{1}{5!} \, \beta_a \, (1-r)^5 + \frac{1}{6!} \, \gamma_a \, (1-r)^6$. The coefficients $\left\{ \alpha_a, \beta_a, \gamma_a \right\}$ are determined by least squares regression. This fitting naturally has large numerical errors, and so the set $A \coloneqq \left\{ (a, \alpha_a) \mid a \in \textrm{grid} \right\}$ must be regarded as a noisy sampling of the fourth derivative $F^{(4)}\big(\frac{\pi\,a}{2}\big)$ evaluated on the boundary $r = 1$. Rather than applying standard noise-reduction filters (e.g. moving averages) on $A$, we can achieve significantly better results if we take into account the fact that $F^{(4)}$ is a smooth function of $\theta$ in the continuum limit. Since we have $\partial_a F^{(4)}\big(\frac{\pi\,a}{2}\big) = 0$ at both $a=0$ and $a=1$, we can expand it spectrally as
\begin{equation}
F^{(4)}\big( {\textstyle \frac{\pi\,a}{2}} \big) = \sum_{n=0}^{\infty} f_n \cos\mathopen{}\left(\mathclose{} n  \, \pi \, a \mathopen{}\right)\mathclose{} \, .
\label{eq:spectral}
\end{equation}
We can therefore obtain a good approximation for $F^{(4)}$ by fitting the coefficients $f_n$ to the first $N$ terms in the series above. Once again, we determine these $f_n$ by applying least squares regression on the set $A$. In this paper, we managed to achieve good results at $N = 20$.

Next, we note that Einstein's equations imply that we should have
\begin{equation}
\varepsilon(a) \coloneqq T^{(4)}({\textstyle \frac{\pi\,a}{2}}) + Y^{(4)}({\textstyle \frac{\pi\,a}{2}}) + U^{(4)}({\textstyle \frac{\pi\,a}{2}}) + V^{(4)}({\textstyle \frac{\pi\,a}{2}}) = 0
\label{eq:tracelessfunc}
\end{equation}
on the $r=1$ boundary. Numerical errors mean that we can never expect the functions obtained by fitting $f_n$ as described above to yield $\varepsilon(a) \equiv 0$ exactly. However, we noticed that in many cases $\varepsilon(a)$ is actually of the same order of magnitude as the $F^{(4)}$ themselves, even though the DeTurck vector norm $\sqrt{\xi^i \xi_i}$ suggests that these solutions should have very small errors. The nonzero $\varepsilon(a)$ therefore seems to contain some systematic discrepancy beyond what one would expect from pure numerical errors.

With this in mind, we manually enforce \eqref{eq:tracelessfunc} by subtracting $\varepsilon(a) / 4$ from each of the functions $\left\{ T^{(4)} , Y^{(4)} , U^{(4)} , V^{(4)} \right\}$. One way to gauge the accuracy of our procedure is to calculate the total energy v.e.v. from the stress tensor,
\begin{equation}
\left\langle E \right\rangle = - 4 \, \pi^2 \, \ell^3 \int_{0}^{\pi/2} \dd\theta \cos\theta \, \sin\theta \, \left\langle T^{t}_{\phantom{t} t}(\theta) \right\rangle .
\end{equation}

To our surprise, this seemingly \emph{ad hoc} procedure resulted in energy densities which, when integrated, agree remarkably well with the black hole masses as calculated by first law method as described in \S\ref{subsec:mass}, with differences ranging between 0.0005\% and 0.1\%. These results were obtainable by following the above procedure exactly, without having to fine-tune it for each particular solution.

We will leave the rigorous analysis of our methods for future work, however we will make a few comments here. The imposition of \eqref{eq:tracelessfunc} amounts to solving the leading-order term of Einstein's equations on the AdS boundary. At the computational level, there are infinitely many ways to do this. One could add unequal proportions of $\varepsilon(a)$ to each function, or apply some completely different operations altogether. Our choice corresponds to pulling out a conformal factor from the boundary metric and imposing Einstein's equations by only modifying this conformal factor. We note that this bears a striking resemblance to the usual conformal decomposition widely used elsewhere in numerical relativity, and it would be interesting to see if a formal justification can be made for its use in this context.

\subsection{Results}

\begin{figure}[t!]
\begin{center}
\begin{subfigure}[lt]{0.488599\linewidth}
    \centering
    \includegraphics[width=7.5cm]{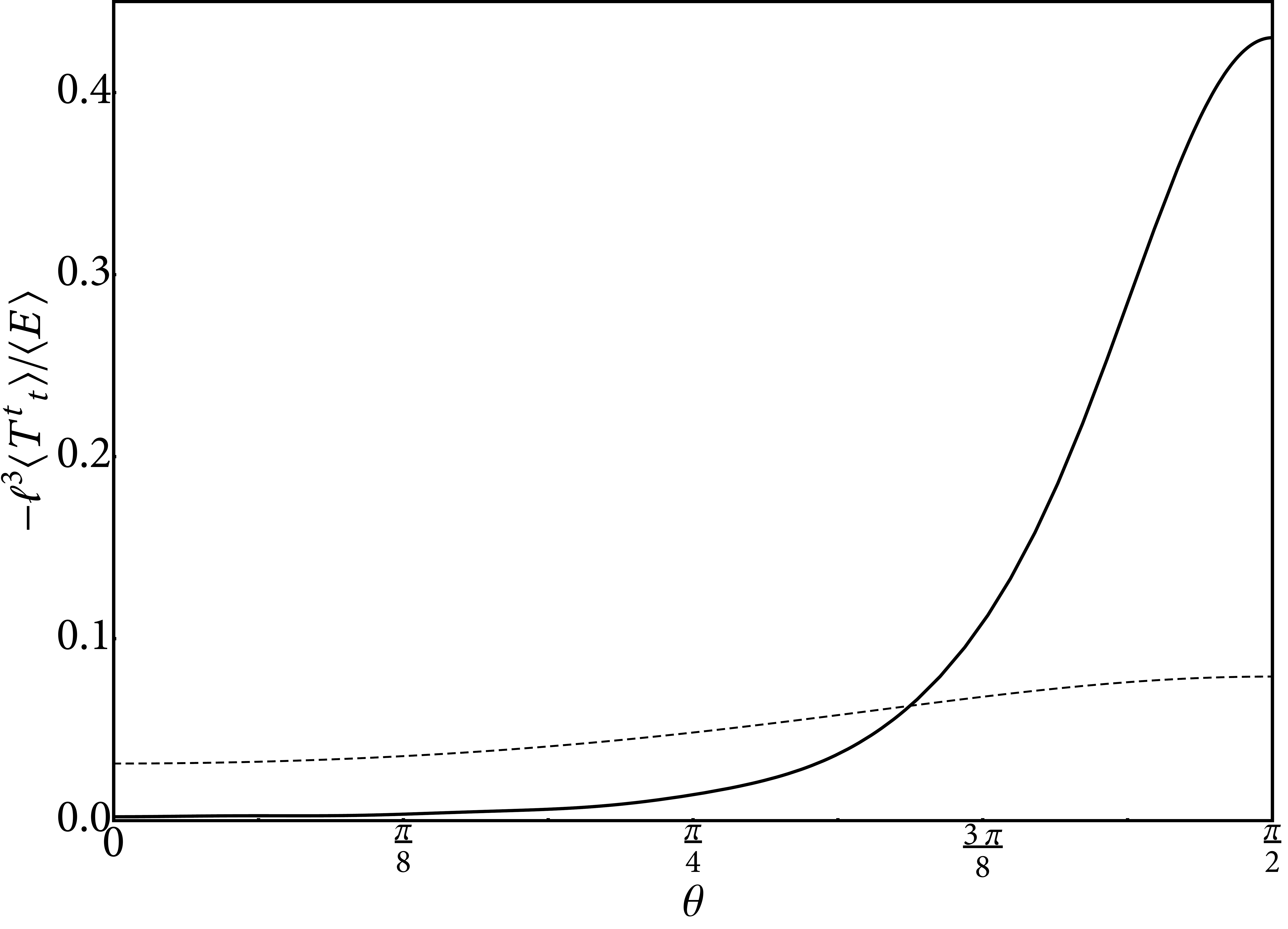}
    \caption{$\kappa\,\ell=2$, $\Omega_H\,\ell=1.07704$}
    \label{subfig:E1}
\end{subfigure}%
\begin{subfigure}[lt]{0.511401\linewidth}
    \centering
    \includegraphics[width=7.8cm]{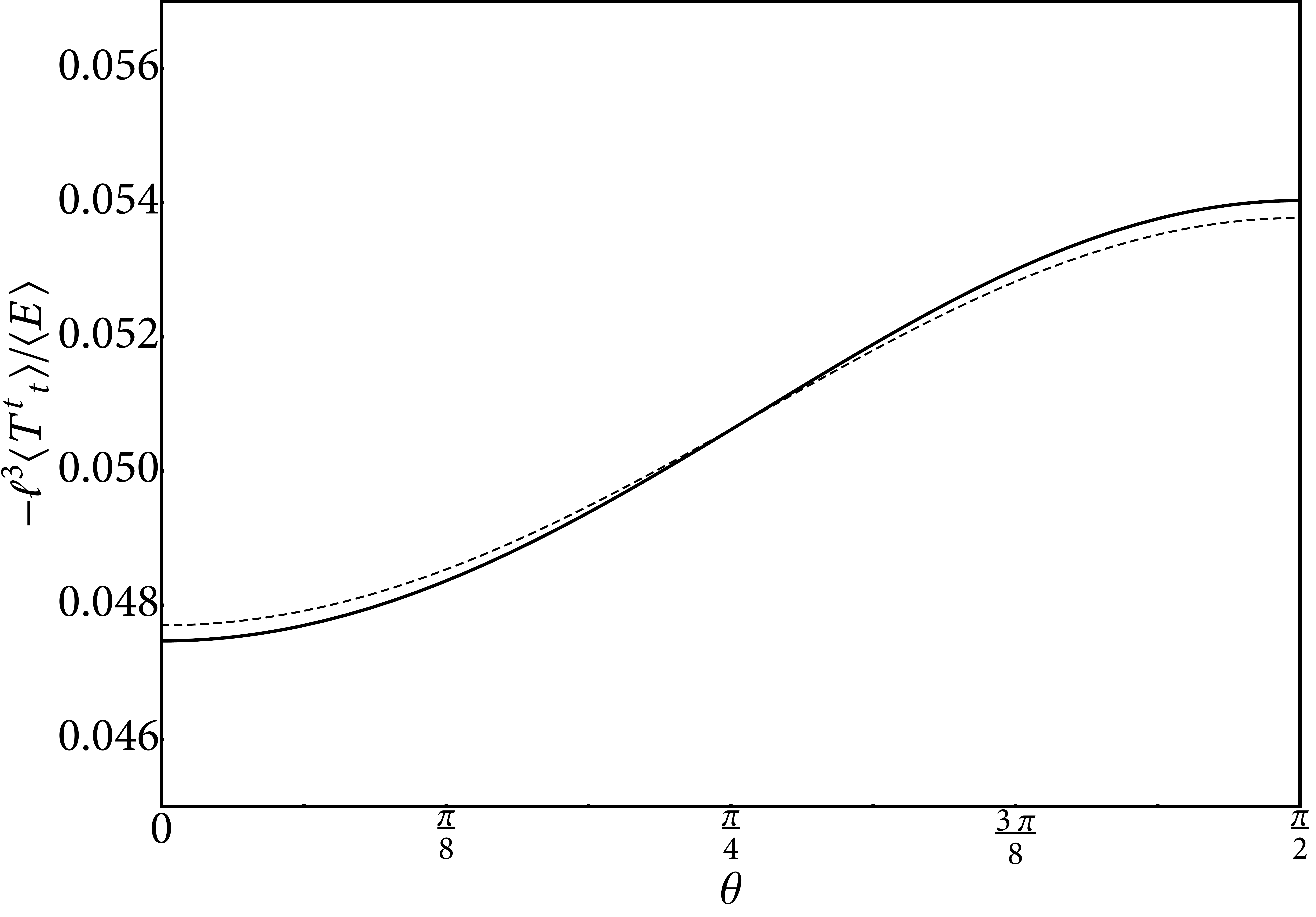}
    \caption{$\kappa\,\ell=2$, $\Omega_H\,\ell=4.53557$}
    \label{subfig:E2}
\end{subfigure}
\begin{subfigure}[rb]{0.5\linewidth}
    \centering
    \includegraphics[width=7.5cm]{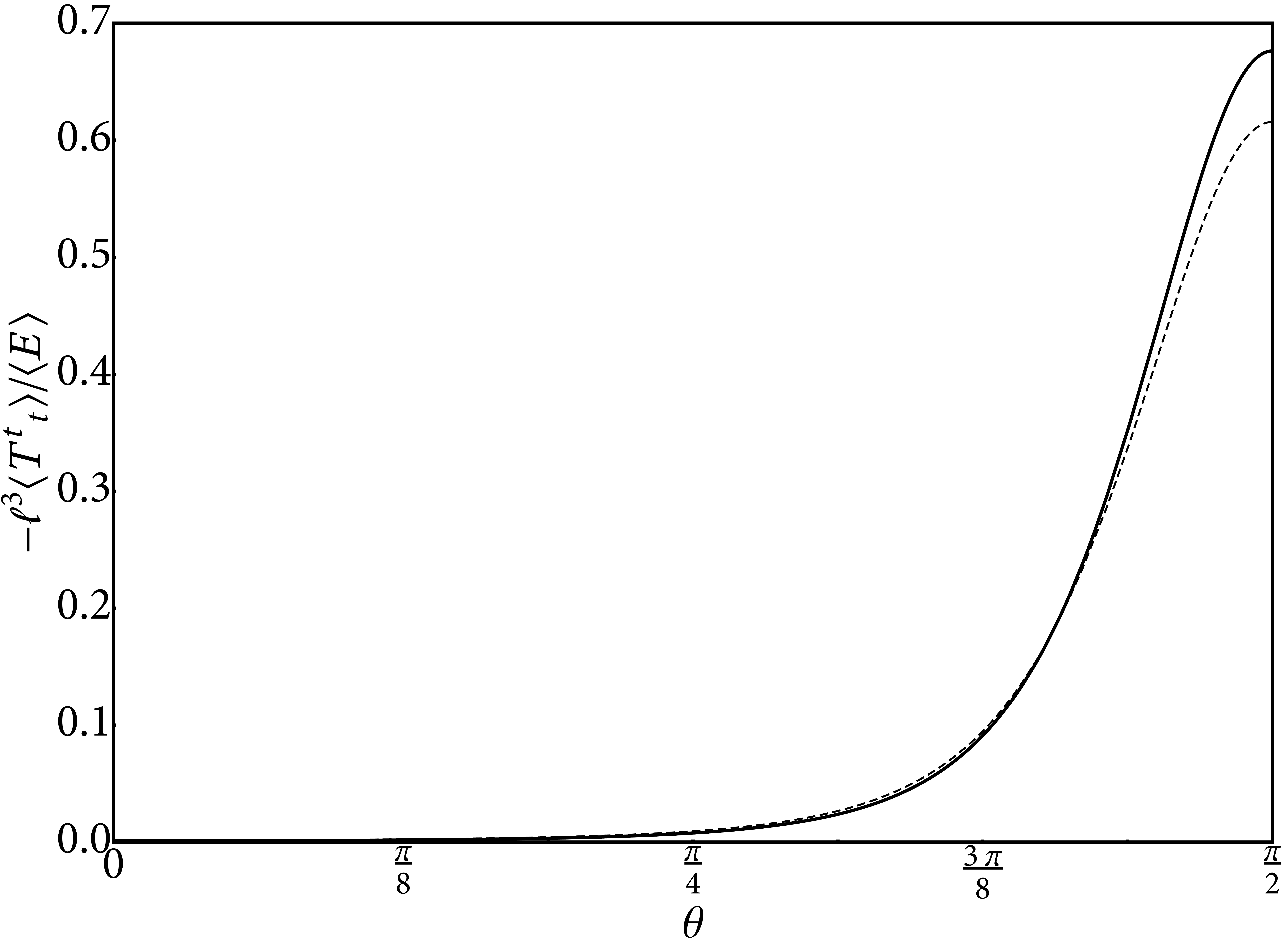}
    \caption{$\kappa\,\ell=1$, $\Omega_H\,\ell=1.05048$}
    \label{subfig:E3}
\end{subfigure}%
\begin{subfigure}[rb]{0.5\linewidth}
    \centering
    \includegraphics[width=7.5cm]{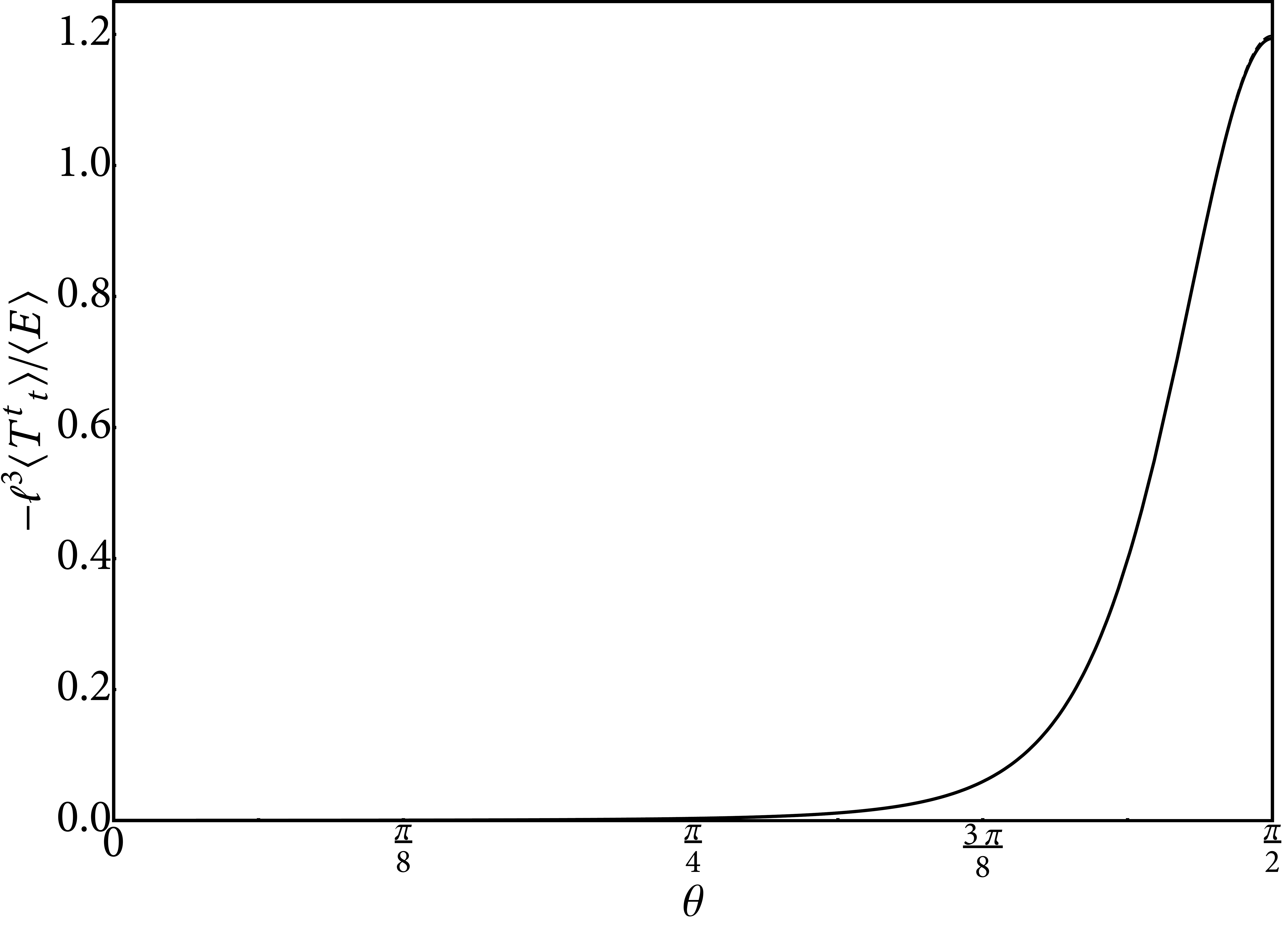}
    \caption{$\kappa\,\ell=0.5$, $\Omega_H\,\ell=1.03750$}
    \label{subfig:E4}
\end{subfigure}
\end{center}
\caption{Energy density distribution, normalised by the total energy of the CFT, as a function of the polar angle $\theta=\frac{\pi\,a}{2}$ on the boundary $S^3$. The data for the black ring is represented as solid black curves, while the data for the corresponding spherical black hole is represented as dashed curves. Firstly, \subref{subfig:E1} shows the energy density distribution for the same long and thin ring as in Fig. \ref{subfig:rad1}. The ring approaches the boundary along the axis of symmetry that goes through one of the poles of the $S^2$ and hence the energy density is concentrated near the corresponding pole of the boundary $S^3$, while the spherical black hole exhibits a much more uniform distribution. Next, \subref{subfig:E3} shows the distribution for a typical fat ring, while \subref{subfig:E2} and \subref{subfig:E4} show the distributions for rings approaching the membrane limit. In all of these cases, the energy densities corresponding to the black ring and the rotating black hole are very similar. Indeed, the two curves become virtually indistinguishable as one moves further towards either of these two limits. }
\label{fig:Edensity}
\end{figure}

In this subsection we present the results for the stress tensor of  CFT states dual to the black rings for some representative configurations. We will concentrate on the energy density distribution on the boundary $S^3$ and shall compare it to the energy density distribution of the rotating AdS black hole at the same temperature and angular velocity, using the results of \cite{Bhattacharyya:2007vs}. 
 
In order to make the correlation between the horizon geometry and the stress tensor of the boundary CFT apparent, in Fig. \ref{fig:Edensity} we have depicted the energy density distribution for same black rings and black holes as in Fig. \ref{fig:Rad}. For thin long rings, Fig. \ref{subfig:E1}, the energy density gets concentrated on one of the poles of the boundary $S^3$, whilst it is negligible on the other pole. The reason is that the ring gets very close to the boundary only along one of the axes of symmetry; the other axis goes through the hole of the ring and hence the energy density in that direction is negligible. Therefore, for  such configurations, it becomes particularly simple to distinguish states dual to black rings from the states dual to spherical black holes.  As the black ring becomes fatter, Figs. \ref{subfig:E2}-\ref{subfig:E4}, the energy densities of the black hole and the black ring approach each other, and the distinction between the two becomes less obvious. This is expected since these two phases should eventually merge. This is particularly striking near the membrane limit, Fig. \ref{subfig:E4}, for which it is very hard to distinguish the energy density corresponding to the ring from that of the black hole. The fact the we see that the energy densities of the two phases approach each other in this limit is reassuring of the correctness of our calculations.  Note that in the membrane limit the energy density also gets concentrated on one of the poles. The reason is that in this limit,  the bulk solution spreads out on the plane of rotation whilst it becomes infinitely thin in the transverse directions. Therefore, the energy density should get very large (eventually diverge) around the pole of the $S^3$ that connects to the rotation plane, and be negligible around the other pole.    
 
We have noted in \S\ref{sec:thermo} that the thermodynamic behaviour of the AdS black rings is qualitatively similar to that of the small rotating black holes in AdS. This gets reflected on the dual stress tensor in the sense that the latter does not fall into the hydrodynamic regime, even though  the stress tensor for rings can be quite different from the stress tensor corresponding to black holes with the same temperature and angular velocity. This is result is unsurprising since \cite{Bhattacharyya:2007vs} solved the relativistic Navier--Stokes equations on $S^3$ for stationary fluid configurations and they only found the solutions corresponding the large rotating black holes in AdS.

%%%%%%%%%%%%%%%%%%%%%%%%%%%%%%%%%%%%%%%%%%%%%%%  
\subsection*{Acknowledgements}
%%%%%%%%%%%%%%%%%%%%%%%%%%%%%%%%%%%%%%%%%%%%%%% 
 
 We would like to thank H. Reall and R. Emparan for discussions. PF and ST are supported by the European Research Council grant ERC-2011-StG279363HiDGR. PF is also supported by the Stephen Hawking Advanced Research Fellowship from the Centre for Theoretical Cosmology, University of Cambridge. The computations presented in this paper were undertaken on the COSMOS Shared Memory system at DAMTP, University of Cambridge operated on behalf of the STFC DiRAC HPC Facility. This equipment is funded by BIS National E-infrastructure capital grant ST/J005673/1 and STFC grants ST/H008586/1, ST/K00333X/1.

%%%%%%%%%%%%%%%%%%%%%%%%%%%%%%%%%%%%%%%%%%%%%%%     
%%%%%%%%%%%%%%%%%%%%%%%%%%%%%%%%%%%%%%%%%%%%%%%  
\bibliographystyle{utphys}
\bibliography{adsrings}

\end{document}